\documentclass[twocolumn]{aastex631}

\usepackage{amssymb}
\usepackage{amsmath}
\usepackage{gensymb}

\begin{document}

\title{Systematic underestimation of polarisation angle dispersion and its consequences for magnetic field strength estimates in star-forming regions}

\author[0000-0001-9751-4603]{Seamus D. Clarke}
\affiliation{Department of Physics, National Cheng Kung University, Tainan, Taiwan}
\affiliation{Academia Sinica, Institute of Astronomy and Astrophysics, Taipei, Taiwan}

\author{Ya-Wen Tang}
\affiliation{Academia Sinica, Institute of Astronomy and Astrophysics, Taipei, Taiwan}

\author{Patrick M. Koch}
\affiliation{Academia Sinica, Institute of Astronomy and Astrophysics, Taipei, Taiwan}

\author{Gary A. Fuller}
\affiliation{Jodrell Bank Centre for Astrophysics, Department of Physics \& Astrononmy, University of Manchester,\\
Oxford Road, Manchester M13 9PL, UK}
\affiliation{I. Physikalisches Institut, University of Cologne, Z\"ulpicher Str. 77, 50937 K\"oln, Germany}

\author{Dawei Xi}
\affiliation{Jodrell Bank Centre for Astrophysics, Department of Physics \& Astrononmy, University of Manchester,\\
Oxford Road, Manchester M13 9PL, UK}



\begin{abstract}
Polarised dust emission observations are a valuable tool to infer the structure of the magnetic field and the dispersion of polarisation position angles may be used to estimate magnetic field strengths. A natural consequence of magneto-dynamic turbulence is for the angular dispersion to have a length-scale dependence, making the measurement of angular dispersion non-trivial. In this paper, we present a study of parametrised, scale dependent maps, focusing on the effect of pixel size and beam convolution on the measured angular dispersion when using the commonly employed unsharp-masking and structure function methods. We find that \textit{in all cases the measured angular dispersion is underestimated} compared to the true value. The degree to which the measured angular dispersion is underestimated varies by factors of 1-10 when measured on scales of 1-3x the beam size, and depends on the underlying structure of the polarisation angle field. This suggests that currently derived magnetic field strengths using angular dispersions are chronically overestimated, potentially leading to an overly magnetically-dominated view of star formation.  We present a method to estimate a correction factor to account for this and apply it to JCMT Orion A OMC-1 observations. We find that the magnetic field in OMC-1 is predominately found to vary on scales much larger than the JCMT's 14'' beam and has a rather low degree of unresolved dispersion, leading to a correction factor of only $\sim$1.6 for angular dispersion measured at a scale of 14''/0.028 pc.
\end{abstract}



\section{Introduction}\label{SEC:INTRO}%
The complexity of the interstellar medium (ISM) arises from multiple physical forces (e.g. gravity, turbulence, radiation magnetic fields) simultaneously interacting on multiple spatial scales \citep[see][for a recent review on the multi-scale ISM]{Pin23}. While magnetic fields have been known to be such a multi-scale presence in the ISM for many decades \citep{Hil49,Mes56}, it is only recently that a large number of sensitive observations across numerous scales has been possible due to the expanded capabilities at a wide range of telescopes \citep[e.g.][]{Koc14,Hul17,Pat17,Kwo18,Tan19,Wan19,Pil20,Lee21,Li22,Wang24}. This has been supported by numerous simulation and theoretical works looking at the role of magnetic fields from molecular cloud scales to protostellar disc scales \citep[e.g.][]{Sei11,Joo12,Che15,Wur16,Ino18,Wur19,Sei20,Kim21,Iba22,Gan23,Gan24}.

A commonly used tracer of the magnetic field in the ISM is the polarised thermal dust emission which is produced due to the short axis of ellipsoidal dust grains being aligned with the magnetic field \citep{DavGre49,Cud82,Hil84,Hil88,LazHoa07,And15}. Such observations of polarised dust emission, using the Stokes Q and U maps, allow the construction of a set of polarisation segments, which when rotated by 90$\degree$, trace the plane of the sky components of the magnetic field. An important quantity derived from such segments is the angular dispersion in their direction as it is may be used to determine magnetic field strengths via the Davis-Chandrasekhar-Fermi method \citep[see][for a recent review on this and other observational methods to study magnetic fields in star forming region]{Liu22}. It is thus important to understand how observational effects and measurement techniques limit our information of this angular dispersion.

\citet{Hei01} previously studied the effects of beam convolution on the measured angular dispersion by using synthetic far-infrared polarised observations from a set of magneto-hydrodynamic simulations. They showed that beam convolution leads to a reduction in the measured angular dispersion compared to the true underlying value, thus leading to overestimates of magnetic field strengths when using the Davis-Chandrasekhar-Fermi method (DCF), or its variants. Further, \citet{Hou09} expanded the angular dispersion structure function method to take into account beam dilution and line-of-sight effects, and so reduce their impact on the estimated magnetic field strengths. 

In this paper we expand upon such works by using easily parametrisable `turbulent' angle maps (i.e. where the angular dispersion is found on multiple size scales linked via a power-law power spectrum) to study how angular dispersion is affected by pixel size and beam convolution, and how this effect appears in common techniques to measure angular dispersion, i.e. the unsharp-masking method \citep{Pat17} and the angular dispersion structure function. Furthermore, we demonstrate how a detailed understanding and quantification of these effects can be used to constrain the properties of the underlying magnetic field structure, and provide a means to apply a correction factor to the unsharp-masking method.

We note that additional factors beyond pixel size and beam convolution affect the polarisation angle as measured from observations, and its link to magnetic field strength, namely the issue of line-of-sight integration and dust alignment. These effects are beyond the scope of this work which aims to focus on the biases/uncertainties from the technical methods of measuring angular dispersion; many prior works have investigated these effects \citep[e.g.][]{MyeGoo91, WieWat04, Fal08, Whi08,Hou09, Hull14, Hoa21}

This paper is structured as follows: in section \ref{SEC:METHOD} we detail the manner in which we generate the `turbulent' angle maps for our numerical experiment; in section \ref{SEC:RES} we present the main results of this paper for different power spectra; in section \ref{SEC:DIS} we discuss the implications of these results and in section \ref{SEC:ORION} show the application of our analysis to JCMT data of Orion A OMC-1; in section \ref{SEC:CON} we conclude.

\begin{figure*}
\centering
\includegraphics[width=0.95\linewidth,trim={1.75cm 2.25cm 0.75cm 3.75cm},clip]{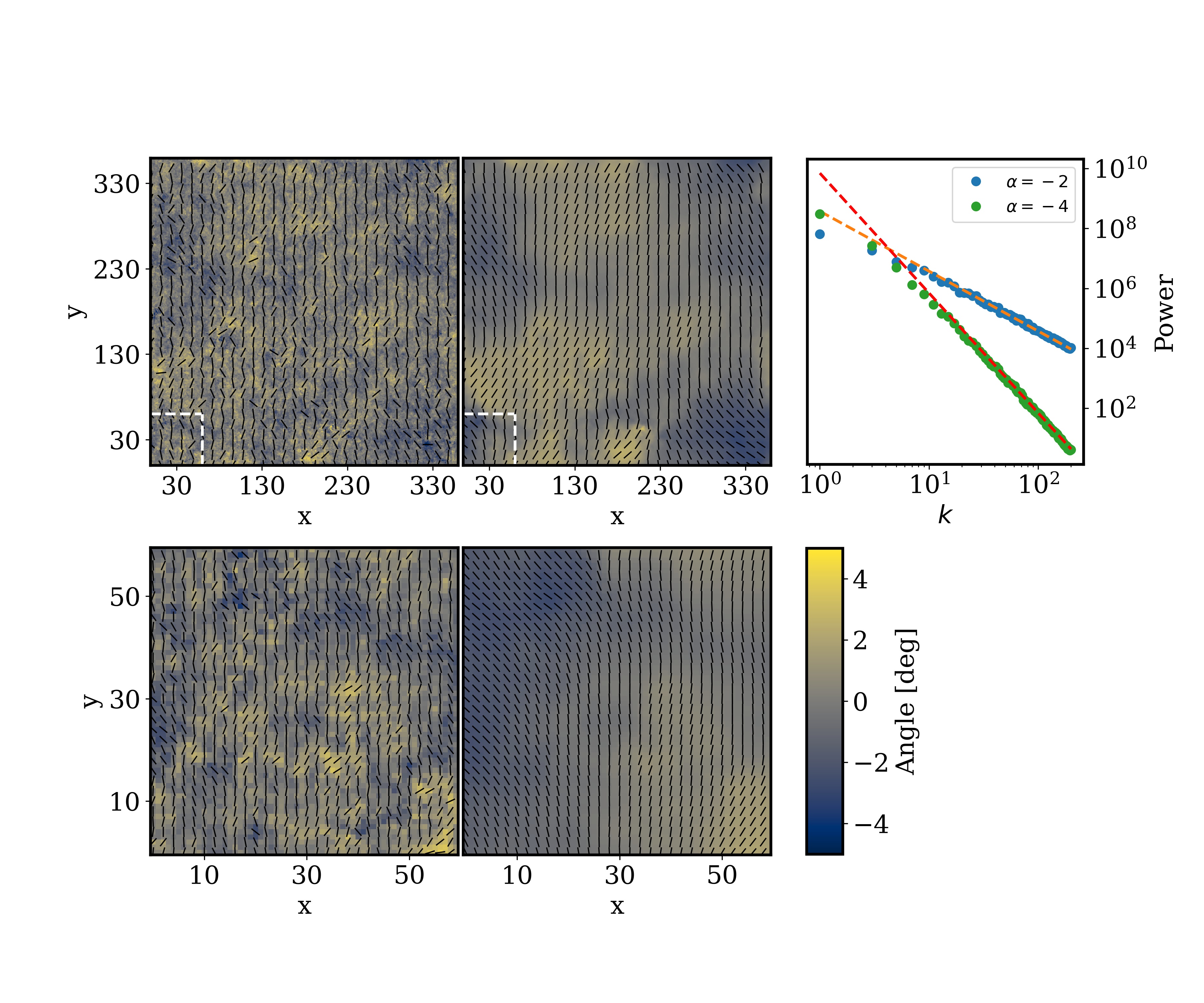}
\caption{(Top left and middle) Two examples of angle maps generated using the method described in section \ref{SEC:METHOD}, where N=360, $k_{\rm{min}} = 1$, and $\alpha=-2$ or $\alpha=-4$ respectively. (Bottom left and middle) Zoom-ins of the region enclosed by the white square in the angle maps above. The black segments on each panel show the angle represented, multipled by a factor of 20 to better illustrate the variation. (Right) The power spectra (circles) calculated from the two example maps, where blue corresponds to $\alpha=-2$ and green to $\alpha=-4$. The orange and red lines also have power-law exponents of -2 and -4 respectively, and show excellent agreement with the power spectra derived from the maps.}
\label{fig::ex_maps}
\end{figure*}  

\section{Generating angular dispersion maps}\label{SEC:METHOD}%
To conduct a numerical experiment one requires a readily parametrised method for producing maps with given angular dispersions. Here we use a method similar to that used to produce turbulent velocity fields (e.g. as described by \citet{Lom15}), which places power in Fourier space according to a power-law distribution, thus allowing control over the spatial modes on which dispersion exists, the proportion of dispersion at larger spatial scales compared to smaller scales, as well as the absolute magnitude of dispersion. There is also a physical motivation for using such a method as a large source of the dispersion in magnetic field segments is thought to come from the coupling of turbulence and the magnetic field \citep{Dav51,ChaFer53,Fed16,Ska21,Laz22}, the two should therefore share similar characteristics. While in reality B-field morphologies are considerably more complex than this single power-law model \citep[for example see the phenomenological models presented in][]{PlanckXII}, a power-law distribution of power in Fourier space for angular dispersion is readily produced by ideal-MHD turbulence without gravity (we show this to be the case for periodic, turbulent box simulations in appendix A).

While the underlying physical quantity of interest is the polarisation angle, $\psi$, this is not directly observed but rather the Stokes Q and U quantities which may be transformed into the angle via:
\begin{equation}
\psi = \frac{1}{2} \arctan{(-U,Q)}.
\label{eq::psi}
\end{equation}
Thus, we use this method to generate the Q and U maps which are then translated into a map of angles. The final result of the method is a 2D image of N$\times$N pixels, each with an angle between -90$\degree$ and 90$\degree$. 

In Fourier space, a map of Q and U may have a power spectrum $P \propto k^\alpha$, where $k$ is the wavenumber corresponding to wavelength $\lambda = L/k$ and $L$ is the side length of the map. When $\alpha$ is more negative, lower wavenumber modes have greater power and greater variation in the quantity is seen on larger physical scales compared to smaller physical scales. When $\alpha = 0$, all wavenumbers have the same power. It is thus via the control of $\alpha$ that one may vary the size scale relationship of the Q and U maps, and the resulting angular dispersion.

To construct a map governed by a given power spectrum, one must first construct an amplitude and phase, $a(\underline{k})$ and $\phi(\underline{k})$, which are functions of the wavenumber vector $\underline{k} = (k_x,k_y)$. The amplitude  is given by:
\begin{equation}
a(\underline{k}) = \sqrt{P(k)} \mathcal{N} = k^{\alpha/2} \mathcal{N},
\end{equation}
where $k = \sqrt{k_x^2 + k_y^2}$ is the magnitude of the wavenumber vector, and $\mathcal{N}$ denotes a random number drawn from a normal distribution with a mean value of 0 and a standard deviation of 1. The phase is given by:
\begin{equation}
\phi(\underline{k}) = 2\pi \mathcal{U},
\end{equation}
where $\mathcal{U}$ denotes a random number drawn from a uniform distribution with a minimum and maximum value of 0 and 1 respectively. This process is performed twice, once to derive the amplitude and phase to construct the Q map, and again for the construction of the U map.

Only vectors corresponding to integer values of $k_x$ and $k_y$ are given non-zero values if they satisfy the following conditions:
\begin{equation*}
0 \leq \; k_y < k_{\rm{max}} / 2,
\end{equation*}
\begin{equation*}
0 \leq \; k_x < k_{\rm{max}} / 2, \; \; \; \mathrm{if} \; k_y = 0,
\end{equation*}
\begin{equation*}
-k_{\rm{max}} / 2 \leq \; k_x < k_{\rm{max}} / 2, \; \; \; \mathrm{if} \; k_y>0,
\end{equation*}
\begin{equation*}
k_{\rm{min}} \leq \; k. 
\end{equation*}
This results in all frequencies having power between $k_{\rm{min}}$ and $k_{\rm{max}}/\sqrt{2}$. For all cases considered here $k_{\rm{max}}$ = N, the number of pixels in each axis of the generated map\footnote{Note that due to this choice the wavelength of the smallest mode is comparable to the pixel size, $\lambda_{\rm{min}} \sim 1.4$. Thus pixel and beam sizes may be read as a multiple of $\lambda_{\rm{min}}$ to better understand the behaviour in general problems.}. The introduction of the parameter $k_{\rm{min}}$ allows one to set a minimum wavenumber (maximum wavelength) for the scale dependence of the angular dispersion below (above) which there exists no power.

\begin{figure*}
\centering
\includegraphics[width=0.9\linewidth,trim={2.6cm 1.5cm 5.5cm 0.5cm},clip]{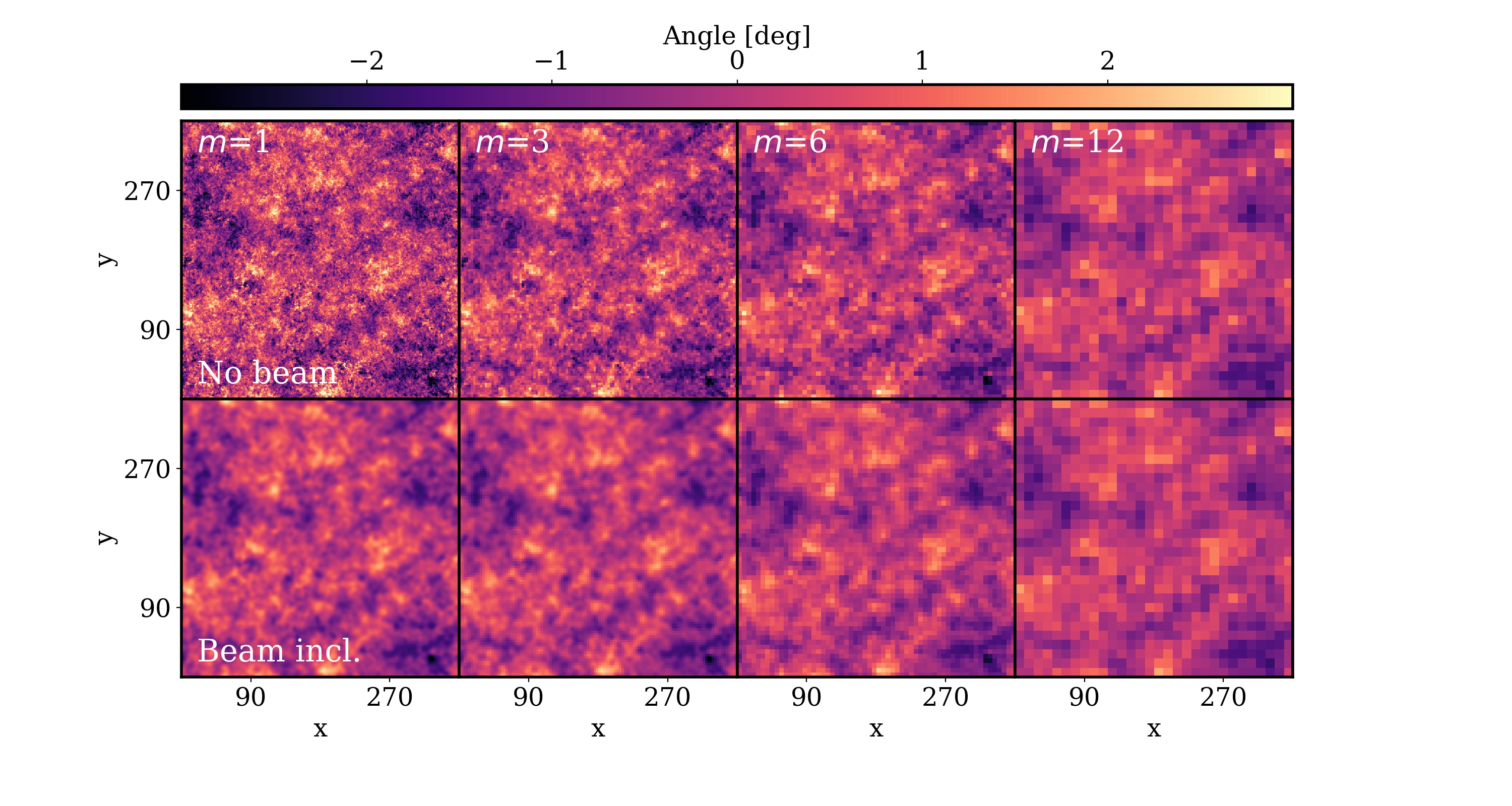}
\caption{(Top row) The angle map resulting from averaging the Q and U maps to larger pixel sizes, $m$, as indicated by the white text. (Bottom row) The same angle map as above but where the Q and U maps have been convolved by a single fixed beam with full-width half-maximum of 6. The Q and U maps are constructed using N=360, $\alpha=-2$, and $k_{\rm{min}}=1$.}
\label{fig::desize}
\end{figure*}  

The Q and U maps are generated by taking the real part of the inverse Fourier transform of the equation $a(\underline{k}) e^{i\phi(\underline{k})}$. As rapid variation between positive and negative values of U when $Q\sim0$ results in sign flipping of large angles (see equation \ref{eq::psi}), we ensure that U is positive everywhere by adding the minimum value of the U map to every pixel. The final U and Q maps may then be transformed into a map of angles, which is then linearly scaled such that the standard deviation is equal to 1$\degree$ and shifted to ensure that the mean is 0$\degree$. The scaling of the mean and standard deviation to these values is arbitrary; setting the mean to 0 results in the largest-scale structure of the resulting maps being vertically aligned, and setting the angular dispersion to 1$\degree$ allows measured dispersion to be read as a fraction of the initial, underlying angular dispersion. The exact choices have no effect on the results in the following sections, with maps having non-zero mean and 30$\degree$ standard deviation being tested. 

It should be noted that as an angle of $-90\degree$ is equal to one of $90\degree$, one must consider this periodicity when calculating the mean and standard deviation of angle distributions. References to a mean or standard deviation will always be to such periodic calculations unless otherwise stated. Here we use the \textsc{Python} functions \textsc{circmean} and \textsc{circstd} provided in the package \textsc{SciPy} \citep{scipy}.  

Two examples of such maps are shown in figure \ref{fig::ex_maps} where N=360, $k_{\rm{min}} = 1$, and $\alpha=-2$ or $\alpha=-4$. Although the two maps share the same angular dispersion, $1\degree$, they are markedly different in their spatial variation. When $\alpha = -4$, quasi-coherent regions can be seen sharing similar angles while this is near absent when $\alpha=-2$ due to the higher relative power at higher wavenumbers resulting in greater small-scale dispersion. This can be better seen when zooming in to a smaller 60x60 region (bottom row of figure \ref{fig::ex_maps}) where very little variation is seen at this smaller scale in the $\alpha = -4$ map, but considerable dispersion is still found for $\alpha=-2$, though clear regions of strong coherence can be found on this small scale. 

To verify that setting $\alpha$ for the Q and U maps results in the same power spectrum slope for the angle map we show the power spectra of the two examples in the right panel of figure \ref{fig::ex_maps}. One sees that there is excellent agreement between the power spectra and the respective power-laws; slight deficits are seen at low $k$ due to these wavenumbers being poorly sampled. 

As the method described above to produce the Stokes Q and U maps uses random numbers, we produce 50 instances of each parameter set ($k_{\rm{min}}$ and $\alpha$) so as to quantify the spread caused by this randomness. 

\subsection{Varying pixel sizes and beam convolution}\label{SSEC:PIXEL}%
Unlike beam convolution, which is unavoidable and determined by the telescope used, pixel sizes are chosen by the observer during imaging. As such it is important to determine how this choice affects the measured angular dispersion, and how it is combined with beam convolution. To investigate this effect on the angular dispersion in our numerical experiment, we take the N$\times$N Stokes Q and U maps with a pixel size of 1, and degrade them to $m$ sized pixels, i.e. a total of N/$m \times$ N/$m$ pixels, where $m = [2,3,4,5,6,8,9,10,12]$. We use N=360 as default to ensure an integer number of pixels for all values of $m$. Degrading is accomplished by finding the mean of $m \times m$ regions on the original map. It is from these Q and U maps with larger pixels that the angle maps are constructed. Figure \ref{fig::desize} shows the effect of larger pixel sizes on a map where N=360, $k_{\rm{min}}=1$ and $\alpha=-2$.

In addition to investigating the effect of pixel size we also include the effect of beam convolution. Before changing the pixel size, the Q and U maps are convolved with a 2D symmetric Gaussian kernel\footnote{An Airy function has also been tested but results in near identical results to a Gaussian kernel.} with a full-width half-maximum of 6. The pixel sizes of these convolved Q and U maps are changed before being converted into angle maps. Thus, our pixel sizes range from 1/6th of a beam size to 2 times the beam size. The bottom row of figure \ref{fig::desize} shows the effect of beam convolution and pixel size together. 

\begin{figure*}
\centering
\includegraphics[width=0.9\linewidth,trim={0.5cm 0cm 4.4cm 2.6cm},clip]{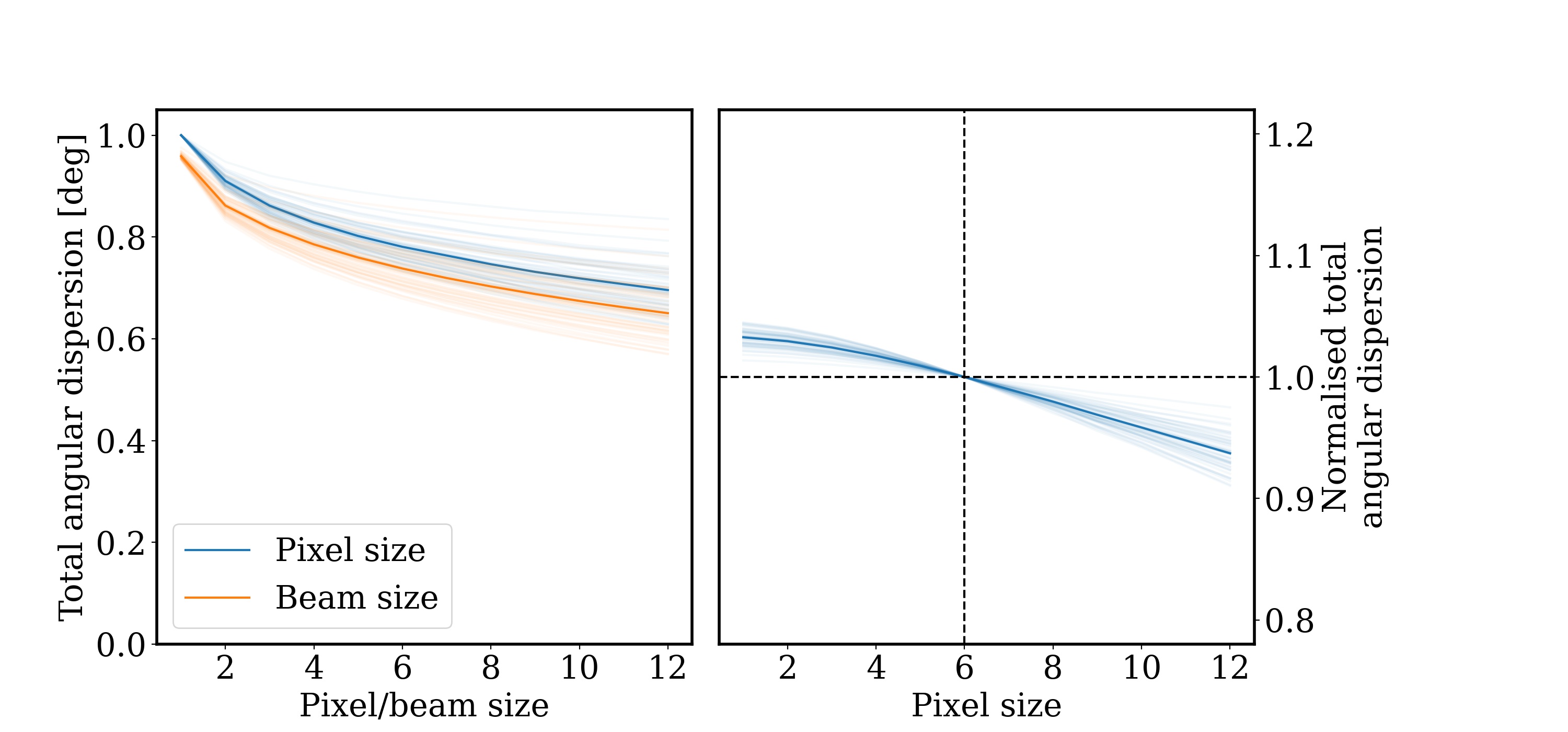}
\caption{(Left) The total angular dispersion as a function of pixel (blue) or beam (orange) size for parameters $\alpha=-2$ and $k_{\rm{min}}=1$. Low opacity lines show the individual results of the 50 realisations of the parameter pair, solid lines show the average value. (Right) The total angular dispersion as a function of pixel size for the same parameters when beam convolution has first been applied, the beam having a full-width half-maximum of 6 (vertical, dashed line). The angular dispersion has been normalised such that when the pixel size is equal to the beam size it has a value of 1.}
\label{fig::beampixel}
\end{figure*}  

We may define a 'median dispersion scale' for the maps we construct such that 50$\%$ of the total dispersion is located below this scale. This is done by integrating these power spectra taking $k_{\rm{min}}=1$. For the $\alpha$ values considered here, -1, -2 and -4, this median dispersion scale is found to be $\sim$ 3, 23, and 260 out of a total map size of 360. Thus with a beam size of 6, these $\alpha$ values range from when the dispersion is predominately found at sub-beam scales ($\alpha=-1$), found at scales from sub-beam to multiple beams ($\alpha=-2$), and found at scales much larger than the beam ($\alpha=-4$); such a range of $\alpha$ may then represent a wide range of possible scenarios in current and future observations. See \citet{Hou11} for the measurement of $\alpha$ in three different star-forming regions.

\section{Results}\label{SEC:RES}%

\subsection{The effects of beam and pixel size on angular dispersion}\label{SSEC:BEAM}%
Figure \ref{fig::beampixel} shows the impact of increasing beam or pixel size on angle maps with the parameters $\alpha=-2$ and $k_{\rm{min}}=1$, i.e. those shown in figure \ref{fig::desize}. One sees that both increasing the beam and the pixel size achieve similar results, that is to reduce the total angular dispersion\footnote{Note that due to the power-law nature of the power spectrum of the dispersion, the behaviour of the total angular dispersion measured over the whole map is the same as the behaviour of a 'local' angular dispersion measured over any smaller region in the map as long as the region size is larger than a few beam/pixel sizes.} in the map with increasing size. This is unsurprising as both processes take the form of local averaging at scales smaller than the pixel or beam size and thus will reduce dispersion at and below those scales. However, it shows a very important point, any measurement of the angular dispersion using a map with resolution, be it beam or pixel size, larger than the smallest turbulent length-scale on which dispersion occurs will always result in an underestimation of the angular dispersion. 

The right panel of figure \ref{fig::beampixel} shows the effect of pixel size on maps which have already been convolved with a beam; here the beam size is fixed to have a full-width half-maximum of 6 and the angular dispersion is normalised such that maps with a pixel size of 6 have a dispersion of 1. Interestingly, it is apparent that pixels with sizes smaller than the beam size slightly alleviate the averaging issue and thus are closer to the true value of the map before beam convolution. This is due to the fact that although pixels within a beam are correlated, spatial variation on scales which are not much smaller than the beam is still encoded. Therefore it is preferable to use pixels smaller than the beam size when measuring angular dispersion, if signal to noise allows, though it does not fully remove the underestimation due to beam convolution, as discussed further in the following sections. Such results have been found in prior works, e.g. \citet{Hei01,WieWat04,Fal08,PatFis19}.

\begin{figure*}
\centering
\includegraphics[width=0.9\linewidth,trim={0.5cm 0cm 4.4cm 2.4cm},clip]{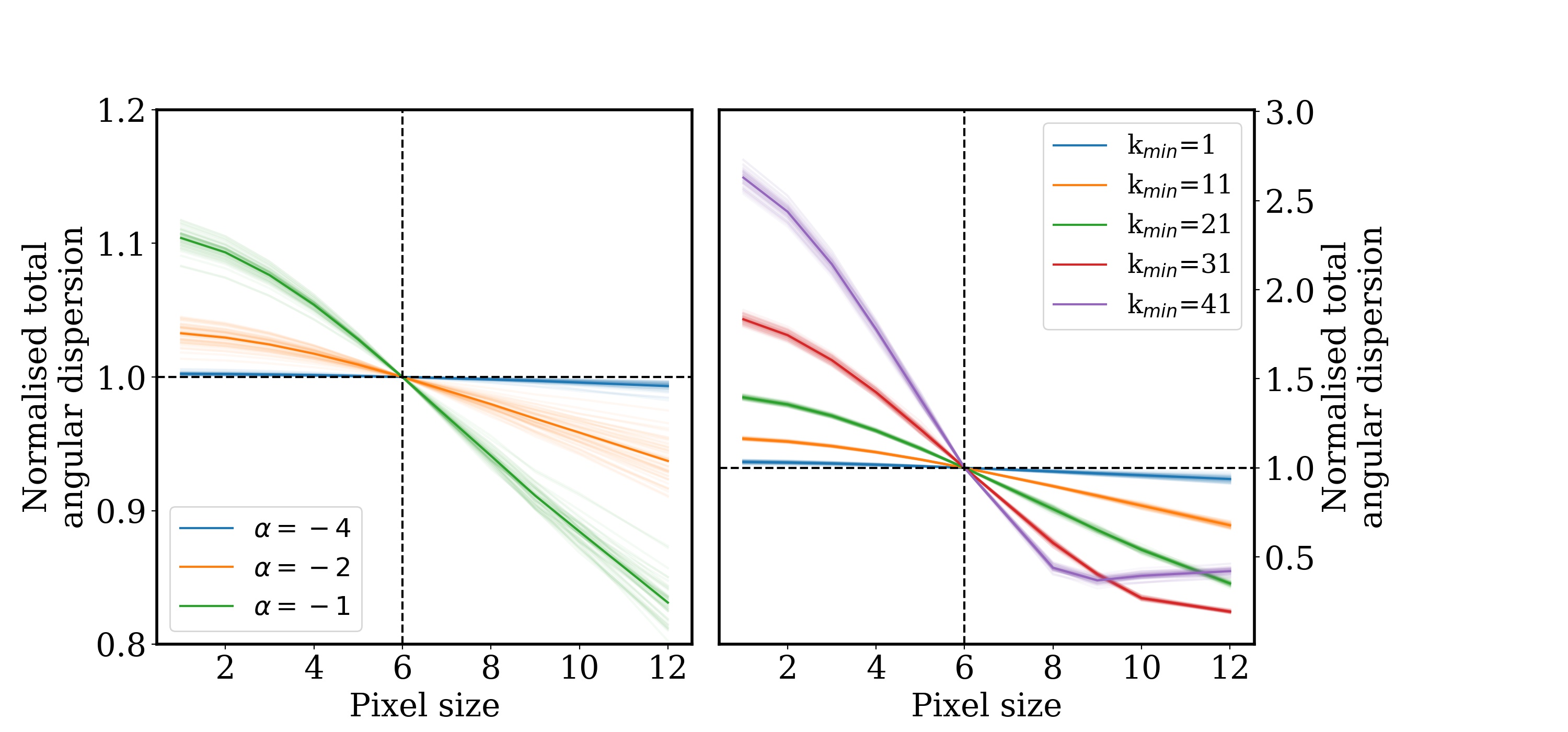}
\caption{(Left) The total angular dispersion as a function of pixel size for maps with a constant $k_{\rm{min}}=1$ and $\alpha$ varies from (green) -1, (orange) -2, and (blue) -4, when beam convolution has first been applied, the beam having a full-width half-maximum of 6 (vertical, dashed line). The angular dispersion has been normalised such that when the pixel size is equal to the beam size it has a value of 1. Low opacity lines show the individual results of the 50 realisations of the parameter pair, solid lines show the average value. (Right) The same as the left panel but where $\alpha$ is kept constant at -2 and $k_{\rm{min}}$ varies from (blue) 1, (orange) 11, (green) 21, (red) 31, and (purple) 41. Note the different scales on the y-axis between the two panels.}
\label{fig::alphakmin}
\end{figure*}  

\subsection{Angular dispersion on the large-scale vs the small-scale}\label{SSEC:ALPHA}%
In the above section, the nature of the angular dispersion was fixed such that $\alpha=-2$ and $k_{\rm{min}}=1$. Here we vary these parameters and therefore investigate how changing the proportion of dispersion found on larger-scales vs smaller-scales impacts the effect of pixels sizes. We use angle maps which have already been convolved with a beam of size 6. As shown above, beam convolution already results in a reduction of angular dispersion compared to the original map without convolution even for pixels of size 1 (right panel of figure \ref{fig::beampixel} for the $\alpha=-2$ case). For $\alpha=-1$ this reduction is 69.7$\%$, for $\alpha=-2$ it is 26.2$\%$, and for $\alpha=-4$ it is 0.7$\%$. Thus, as one would expect, regions where dispersion is predominately found on scales below the beam size, (e.g. $\alpha=-1$), suffer from a much higher reduction of angular dispersion due to beam convolution compared to those where dispersion is predominately found on scales much greater than the beam ($\alpha=-4$).

The left panel of figure \ref{fig::alphakmin} shows the measured angular dispersion as a function of pixel size for maps with a constant $k_{\rm{min}}=1$ and $\alpha=-1,-2,-4$. In all cases one sees that using sub-beam sized pixel results in measuring a larger angular dispersion, i.e. closer to the true value without beam convolution, although still being a factor of 1-3 below the true value. As one might expect, this effect is more pronounced for dispersion maps which have a larger share of dispersion on sub-beam scales ($\alpha=-1$). This suggests that one might be able to use the behaviour of the measured angular dispersion as a function of pixel size to infer the relative share of dispersion of small-scales compared to larger-scales in a qualitative manner.

The right panel of figure \ref{fig::alphakmin} shows the measured angular dispersion as a function of pixel size for maps with a constant $\alpha=-2$ and $k_{\rm{min}}=1,11,21,31,41$. Similar to the variations in $\alpha$, one sees that in all cases larger pixels reduce the measured angular dispersion while sub-beam sized pixels recover a greater share of the underlying dispersion. Also similar to the $\alpha$ variations, maps in which dispersion is predominately found on smaller scales (i.e. larger values of $k_{\rm{min}}$) show a stronger impact on pixel size; however, note the much larger increase in change in the measured angular dispersion, with some maps showing variations of greater than a factor of 2. 

An interesting feature in the right panel of figure \ref{fig::alphakmin}, which is not apparent in the left panel corresponding to variations in $\alpha$, is the turning point in angular dispersion seen at $m=9$ for $k_{\rm{min}}=41$. The location of this turning point corresponds to the associated maximum length-scale for this $k_{\rm{min}}$, $360/41 = 8.78$. A decelerating decline in the angular dispersion for $k_{\rm{min}}=31$ between $m=10$ and $m=12$ is likely also connected to its maximum length-scale, $360/31 = 11.6$, and another turning point may become apparent if larger sized pixels were considered. This result brings up the counter-intuitive possibility that measured angular dispersion may actually increase slightly if larger pixel sizes are considered, but only for those regions in which the dominant large-scale dispersion scale is comparable to the pixel sizes. 

\begin{figure*}
\centering
\includegraphics[width=0.9\linewidth,trim={0.5cm 0.5cm 3.65cm 2.25cm},clip]{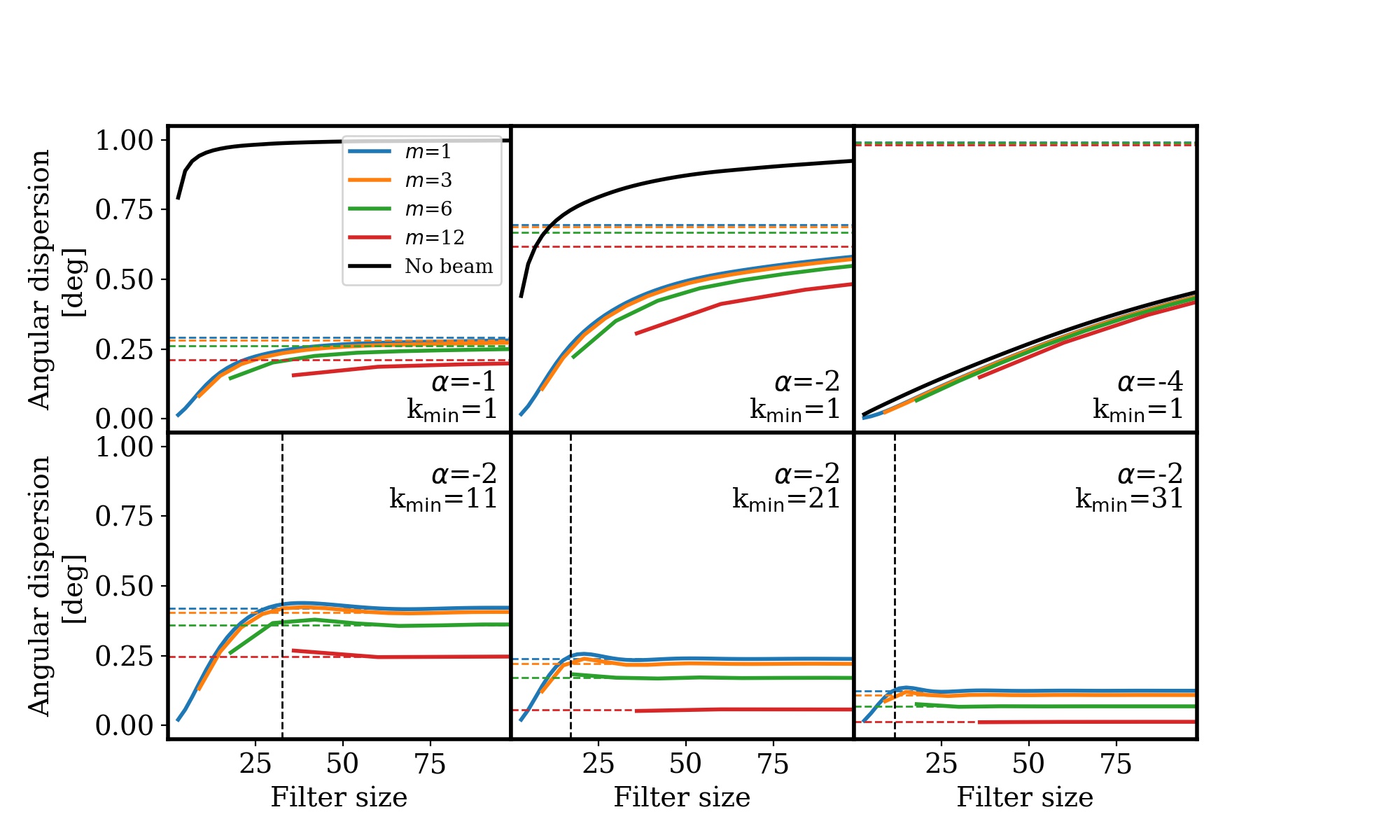}
\caption{Measured angular dispersion using the unsharp-masking method as a function of filter size (i.e. the side length of the square filter) for maps with varying values of $\alpha$ (top row) and of $k_{\rm{min}}$ (bottom row). The measurements using pixels of size 1 (blue), 3 (orange), 6 (green), and 12 (red) are shown for each panel and the maps have been convoluted with a beam of size 6. The total dispersion in the map measured is shown as a horizontal dashed line of the corresponding colour. The black, solid line in the top row shows the same measurement for the corresponding map without beam convolution and $m=1$, i.e. the map without loss of information. In the bottom row, the vertical, dashed, black lines show the size scale associated with the map's value of $k_{\rm{min}}$. Note that only one realisation out of the total of 50 is shown for each parameter pair.}
\label{fig::filtersize}
\end{figure*}  

\section{Consequences for common techniques of measuring the angular dispersion}\label{SEC:COMMON}%

\subsection{Measurements using the unsharp-masking method}\label{SSEC:FILTER}%
When measuring the angular dispersion, it is common to attempt to remove dispersion found on larger scales so as to only measure small-scale `turbulent' dispersion to be used in the DCF method. It should be noted though that as shown in appendix A, turbulence produces dispersion across many scales and as such there is no single length-scale in the B-field morphology at which the dispersion is turbulent; they all are. However, techniques to remove larger scale dispersion do allow one to measure angular dispersion at a specific scale which may be of interest. 

One such technique is the unsharp-masking method \citep{Pat17}. This technique works by considering a square filter of some side length, $L$, centred on a pixel and determining the mean polarisation angle of all the pixels within that square. By subtracting the mean angle within the filter from that central pixel, and repeating this process for all pixels, one is left with a map of residual angles around the local mean. Calculating the standard deviation of this residual map then results in the angular dispersion at scales below the square filter size $L$. Recently this technique has been modified by instead of calculating the local mean within the filter, one may calculate the dispersion within the filter and assigning its value to the central pixel \citep[e.g.][]{Hwa21,Ngoc23}. This results in a map of local dispersion below size scale $L$, instead of just a single number characterising a whole map, and thus allows investigation of the spatial distribution of the B-field angular dispersion. 

The filter size is not fixed but may be chosen to measure dispersion on any specific scale; while this choice is arbitrary, it has typically been set to 3 or 5 pixels to focus on the smallest scales. In this section we investigate the results of varying the filter size, as well as the effect of pixel sizes, to explore how sensitive the measured dispersion is on these choices. 

Figure \ref{fig::filtersize} shows the measured angular dispersion as a function of filter size for beam convolved maps characterised by varying values of $\alpha$ (top row) and of $k_{\rm{min}}$ (bottom row). Each panel shows only 1 realisation out of the total of 50 for each parameter pair for clarity, though there is little difference between each realisation. It is clear across all maps that the measured angular dispersion is highly dependent on the filter size, showing that the choice of it, e.g. 3 or 5 pixels, and the limitations of the observations, i.e. beam size, will have a large impact on the measurement and hence the B-field strength. Due to this, it is important to report the filter size in combination with the measured dispersion as the two are inextricably linked. 

There is clear variation in the dependence on filter size for different values of $\alpha$, with those maps with a greater degree of dispersion on smaller scales quickly reaching a plateau as the filter size increase ($\alpha=-1$) which is not present in the case for maps with a greater degree of dispersion on larger scales ($\alpha=-4$). Such variation suggests that the application of the unsharp-masking method with varying filter sizes, hereafter termed filter-size plots, could be used to characterise the nature of a map of polarisation segments, and thus allow comparisons of B-field morphologies between regions.

For maps with varying $k_{\rm{min}}$ a different feature appears. When the filter size is roughly equal to the scale associated with $k_{\rm{min}}$, i.e. the scale which has the most dispersion, there exists a slight maximum in the measured dispersion. This would make the use of varying filter sizes an efficient method to detect the presence of important scales in the B-field.

Across all map examples, the effect of increasing the pixel size is to decrease the measured angular dispersion at a given scale, and this decrease is greater for maps which have a larger degree of dispersion on smaller scales. This is the same as when measuring the total dispersion across the map. Thus, it is advisable to use sub-beam sized pixels to recover more angular dispersion when applying the unsharp-masking method. 

\begin{figure}
\centering
\includegraphics[width=0.95\linewidth,trim={0.5cm 0.2cm 0.5cm 0.5cm},clip]{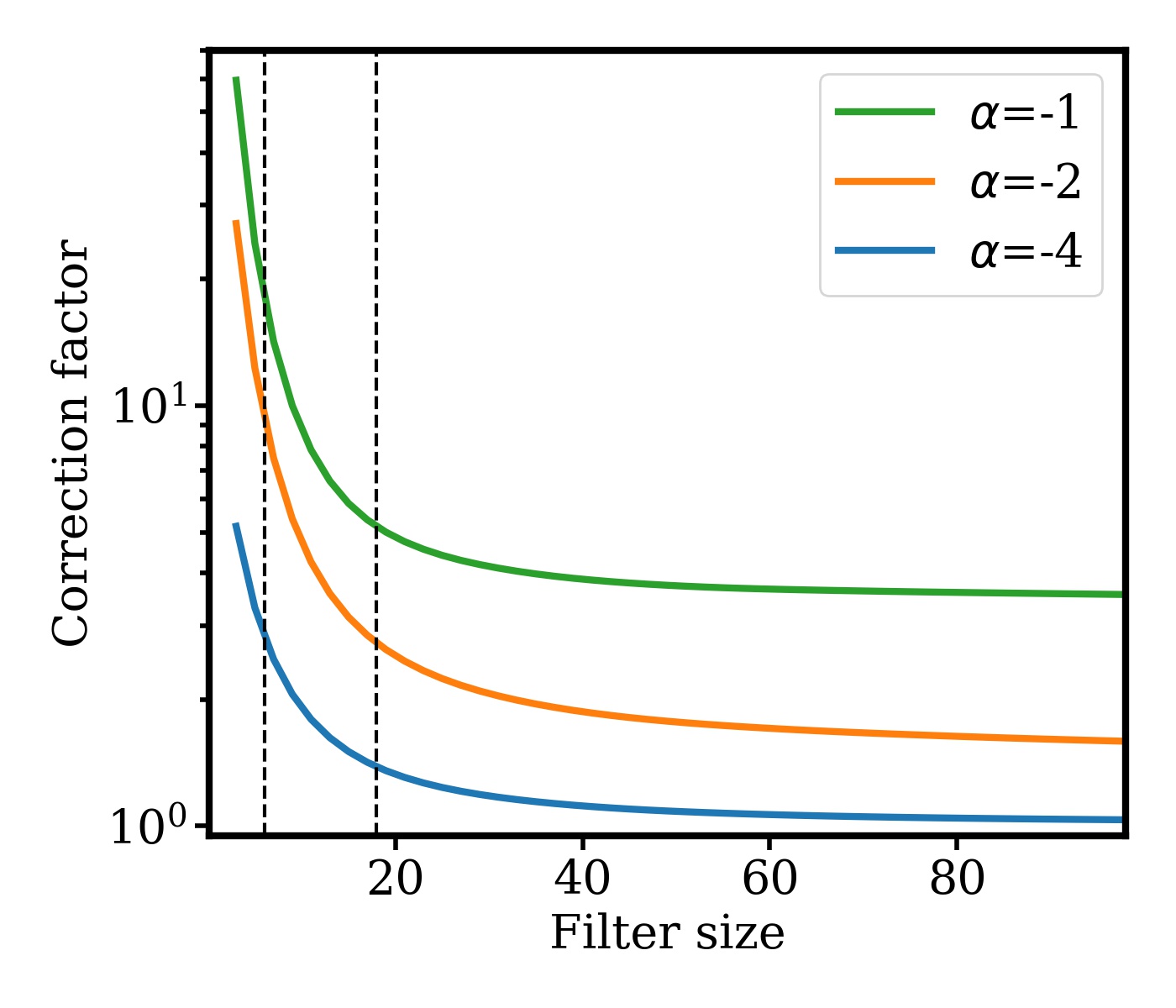}
\caption{The angular dispersion correction factor as a function of the filter size with $\alpha=-1, -2, -4$ (green, orange and blue respectively). The correction factor is defined as $\sigma_{\psi}(l)/\sigma_{\psi,b}(l)$, where $\sigma_\psi (l)$ is the angular dispersion measured at filter size $l$, and the $b$ subscript denotes the measurement is from a beam convolved map. Here the measurements are made for maps with pixel size 1 and $k_{\rm{min}}$=1, and the beam has a size of 6. The two vertical, dashed lines identify when the filter size is equal to the beam size, and three times the beam size.}
\label{fig::filter_correction}
\end{figure}

Considering the effect of the beam convolution on the angular dispersion across filter sizes, the top row of figure \ref{fig::filtersize} shows that the drop in measured angular dispersion compared to the same map without beam convolution is not uniform across filter sizes. Rather, as might be expected, the smaller spatial scales close to the beam size see the larger drop in measured angular dispersion compared to scales much larger than the beam size. To investigate this, we construct a `correction' factor by which the measured angular dispersion from a beam convolved map can be brought into alignment with the angular dispersion at the same filter size without beam convolution. Figure \ref{fig::filter_correction} shows this correction factor for the $\alpha$ values of -1, -2 and -4 for $m=1$ sized pixels.

The correction factor has a strong dependence on the properties on the angular dispersion field, with maps having more dispersion on scales much greater than the beam size ($\alpha=-4$) having considerably smaller correction factors than when the inverse is true ($\alpha=-1$). For filter sizes equal to 1 beam size (here 6), the correction factor is 17.9, 9.2, 2.8 for $\alpha=-1$, -2 and -4 respectively; and for filter sizes equal to 3 beam sizes, it is 5.2, 2.7, 1.4. Thus, it is possible that angular dispersion measured close to the beam size, as is commonly done in observational works, may be underestimating angular dispersion on those scales by factors of 1-10. The exact value of the correction factor is dependent on the properties of the angular dispersion field itself (e.g. $\alpha$, $k_{\rm{min}}$, total angular dispersion, and the beam size relative to $k_{\rm{max}}$) but the shape of the filter-size plot encodes this information and thus would allow an estimate of the correction factor - this will be discussed in more detail in section \ref{SEC:ORION} when applying the technique to real observational data.

We note here that there exist many other proposed correction factors in the field related to estimates of magnetic field strength via the DCF method \citep[e.g.][]{Ost01,Pad01,Fal08,Liu21}. Such works have typically used magnetohydrodynamical simulations in combination with synthetic emission/observations to construct a correction factor which brings DCF-derived magnetic field strengths of such simulated clouds/regions into alignment with the true value. The correction factor introduced here is not related to the DCF method and instead aims to solely correct the measurement of the angular dispersion using the unsharp-masking method; it is thus a factor derived from the observational data itself (i.e. from the shape of the filter-size plot) and which will differ from region to region.

\subsection{Measurements using the angular dispersion structure function}\label{SSEC:STRUCT}%
Another common method of investigating the dispersion found in polarisation segments is the use of a structure function \citep[e.g.][]{Hil09,Hou09,Hou11,Hou13,Hou16}. By quantifying the angular dispersion on multiple spatial scales and fitting an analytically derived equation, one may attempt to separate the small-scale 'turbulent' dispersion from the large-scale 'ordered' field. We apply this technique to the angle maps characterised by various $\alpha$ and $k_{\rm{min}}$ values, and different pixel sizes to investigate their effects on the resulting measured angular dispersion. 

\begin{figure*}
\centering
\includegraphics[width=0.9\linewidth,trim={1.25cm 0.5cm 3.65cm 2.25cm},clip]{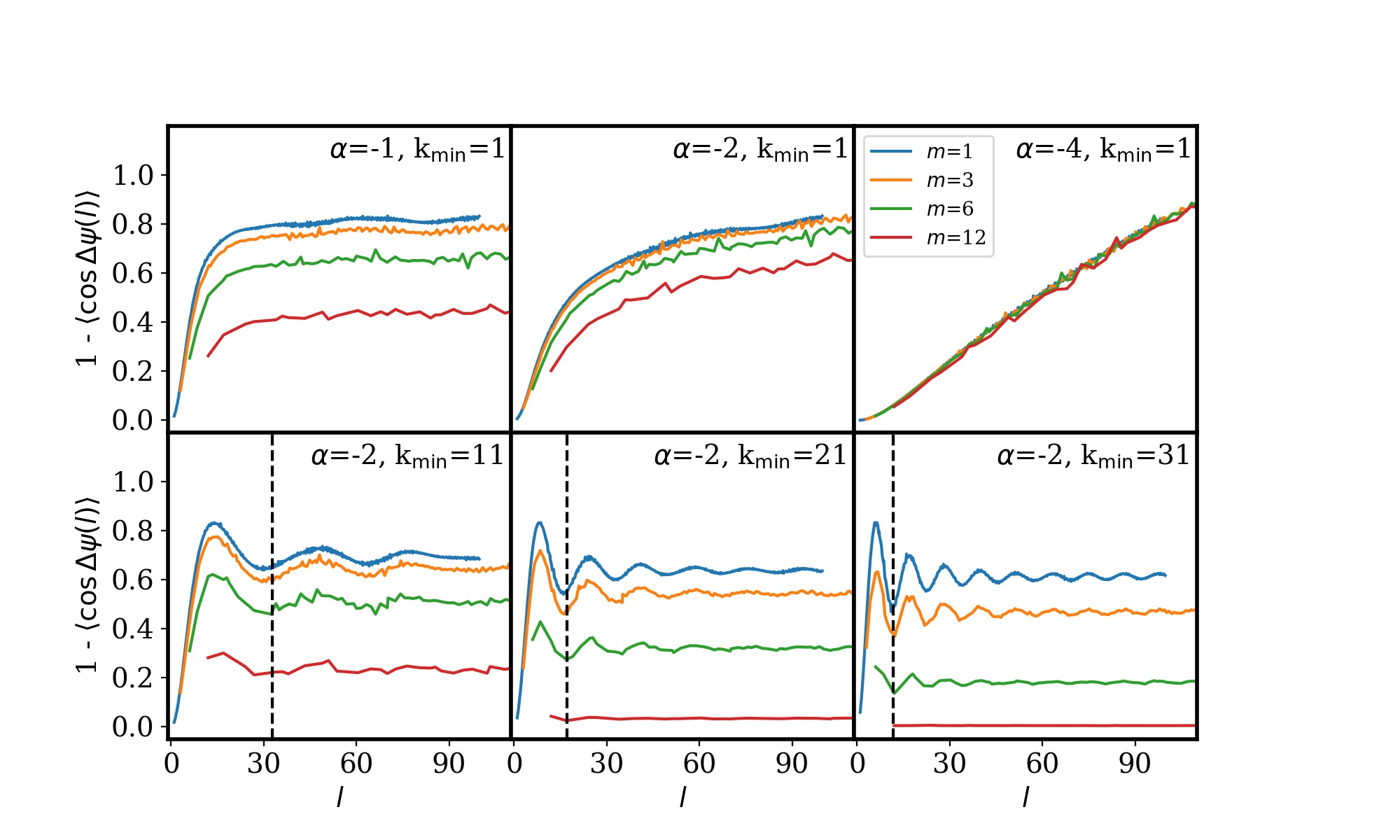}
\caption{Structure functions for maps with varying values of $\alpha$ (top row) and of $k_{\rm{min}}$ (bottom row). The measurements using pixels of size 1 (blue), 3 (orange), 6 (green), and 12 (red) are shown for each panel. In the bottom row, the vertical, dashed, black lines show the size scale associated with the map's value of $k_{\rm{min}}$. Note that only one realisation out of the total of 50 is shown for each parameter pair, that all maps have been convolved with a beam of size 6, and that the y-axis has been normalised so that all 6 panels may be shown at once.}
\label{fig::struct}
\end{figure*}  

Figure \ref{fig::struct} shows the measured structure function for beam convolved maps with varying $\alpha$ (top row) and $k_{\rm{min}}$ (bottom row). Each panel shows only 1 realisation for each parameter pair for clarity, and the structure function has been averaged using bin sizes of 0.25 pixels. One sees numerous similarities with the results of the filter-size plots using the unsharp-masking method, that is, maps whose morphology have considerable dispersion on small-scales reach plateaus relatively quickly but also show the largest dependence on the pixel size used. Secondly, the presence of a maximum dispersion scale, set here by various $k_{\rm{min}}$, leads to a clear oscillatory signature in which the first trough aligns with this spatial scale. The similarity between the two methods is unsurprising as both measure angular dispersion across multiple scales.

Considering the effects of beam convolution \citep{Hou09}, the structure function can be written in the following manner:
\begin{equation}
1 - \langle\cos{\Delta\psi(l)}\rangle = \frac{1}{N}\frac{\langle B_t^2\rangle}{\langle B_o^2 \rangle} \left(1 - e^{-l^2 /2(\delta^2 + 2W^2)}\right) + al^2,
\label{eq::struct}
\end{equation}
where $\Delta\psi(l)$ is the angle difference between two polarisation/B-field segments separated by a distance $l$, $B_t$ and $B_o$ are the turbulent and large-scale, ordered magnetic field strengths, $\delta$ is the turbulent correlation length associated with $B_t$, $W$ is the radius of the beam, $a$ is a coefficient to approximate the large-scale magnetic field structure, and $N$ is the number of turbulent cells along the line-of-sight within one beam. The angled brackets denote the arithmetic mean of the quantity. It is by fitting the derived structure function by this equation with the free parameters $\delta$, $a$ and $\frac{\langle B_t^2\rangle}{\langle B_o^2 \rangle}$, that the plane-of-the-sky B-field strength may be estimated. Higher order terms of the form $\Sigma_k a_k l^{2k}$, may be added to equation \ref{eq::struct} to better model the large-scale field. In this work terms up $l^6$ has been considered but no significant difference is found compared to only including the $l^2$ term as shown in equation \ref{eq::struct}.

The number of turbulent cells along the line-of-sight is typically given by $N = L\frac{\delta^2 + 2W^2}{\sqrt{2\pi}\delta^3}$, where $L$ is the thickness of the structure along the line-of-sight, as this takes into account integration along the line-of-sight \citep{Hou09}. However, for fitting here, as the maps have not been constructed with any such line-of-sight integration we use the form $N = \frac{\delta^2 + 2W^2}{\delta^2}$ \citep{Hou09}. Furthermore, the application to equation \ref{eq::struct} implies that $N \langle B_o^2 \rangle \gg \langle B_t^2\rangle$; as seen from the fitted values of $\frac{\langle B_t^2\rangle}{\langle B_o^2 \rangle}$ and $\delta$ below, this condition is true for the polarisation angle maps considered here. 

\begin{figure*}
\centering
\includegraphics[width=0.98\linewidth,trim={1.5cm 0cm 0cm 0cm},clip]{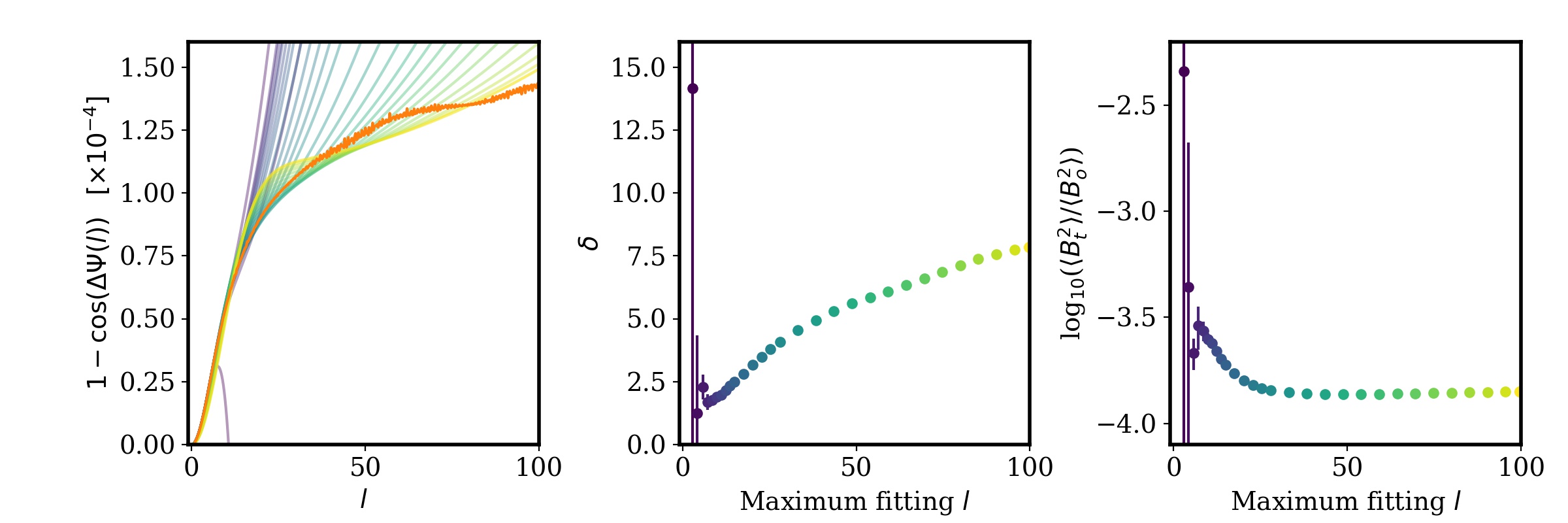}
\caption{(Left) The structure function for a realisation with $\alpha=-2$ and $k_{\rm{min}}=1$ (orange) with the various fits colour-coded by the maximum value of $l$ used. (Middle and right) The fitting parameters $\delta$ and $\frac{\langle B_t^2\rangle}{\langle B_o^2 \rangle}$ respectively as a function of the maximum $l$ included in the fitting, colour-coded in the same manner. The error-bars show the 1$\sigma$ uncertainties from the fitting results.}
\label{fig::struct_fit}
\end{figure*}  

\begin{figure}
\centering
\includegraphics[width=0.98\linewidth,trim={0cm 1.2cm 2.cm 1cm},clip]{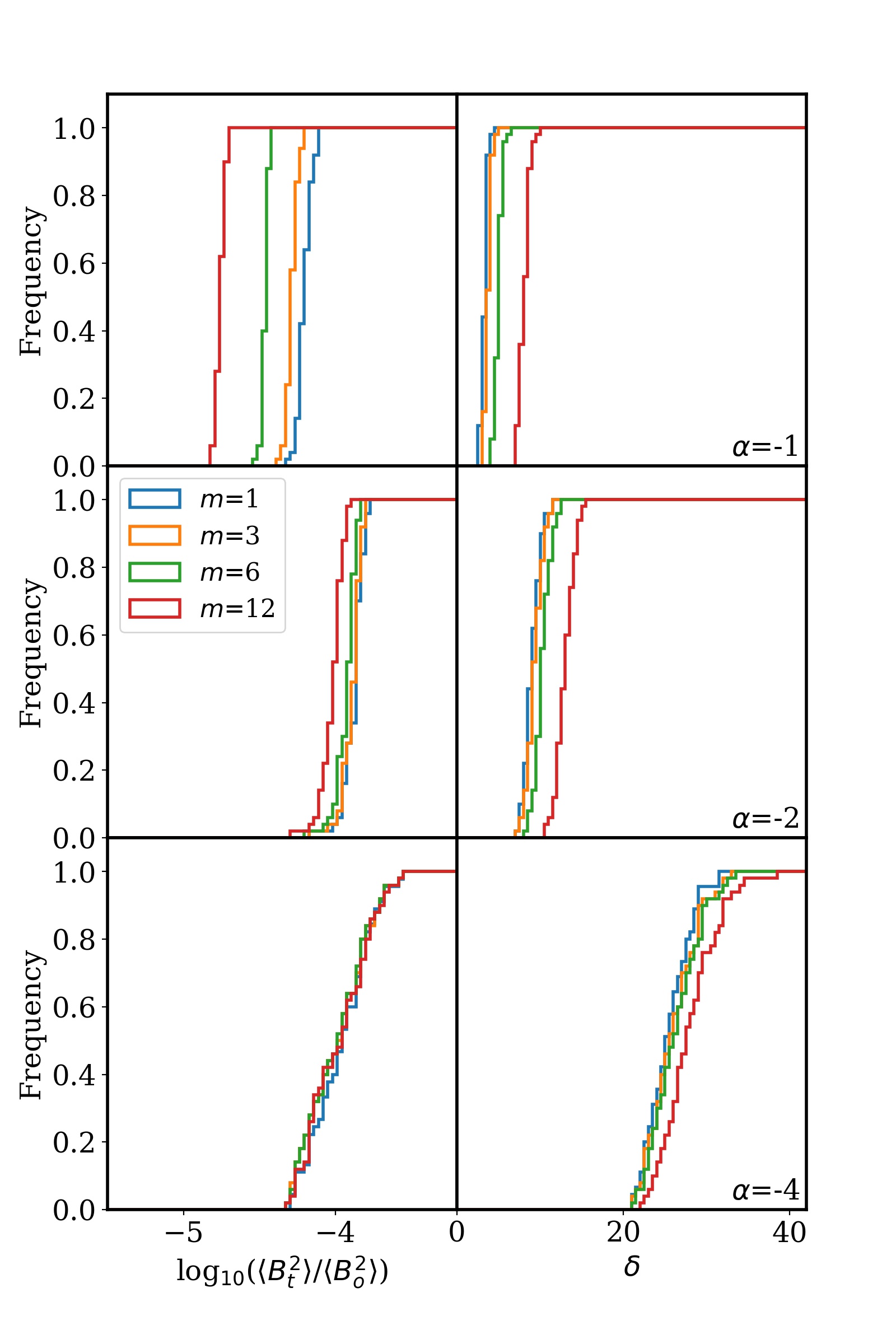}
\caption{Cumulative distribution plots of the fitting results for (left) $\frac{\langle B_t^2\rangle}{\langle B_o^2 \rangle}$ and (right) $\delta$ for the structure functions of maps with $k_{\rm{min}}=1$ and $\alpha$ set to -1, -2 and -4 for the top, middle and bottom rows respectively. The colours indicate the pixel size in the map with $m$=1, 3, 6, 12 shown as blue, orange, green and red lines respectively.}
\label{fig::struct_pixel}
\end{figure}  

One common issue when fitting the structure function is to determine the maximum distance $l$ out to which the function will be fitted. This may be due to the relatively small extent of strong polarisation signal leading to poor sampling at larger distances. To examine this effect, we may vary the maximum distance used when fitting these perfectly sampled structure functions.

Figure \ref{fig::struct_fit} shows the structure function for one realisation with $\alpha=-2$ and $k_{\rm{min}}=1$ with the various fits colour-coded by the maximum value of $l$ used, and their corresponding fitted values for $\delta$ and $\frac{\langle B_t^2\rangle}{\langle B_o^2 \rangle}$. One can see that below l$\sim$30-40, the fitting parameters are highly dependent on the maximum extent of the structure function used, with $\frac{\langle B_t^2\rangle}{\langle B_o^2 \rangle}$ varying by nearly an order of magnitude. Above this value, the fitting parameters are much more weakly dependent on the maximum extent used for fitting. This transition point corresponds to roughly the location in which the structure function begins to plateau into a slower growth. This is not unique to this realisation, or to these particular values of $\alpha$ or $k_{\rm{min}}$, but appears general. For cases like $\alpha=-4$, where the structure function does not possess a plateau, fitting parameters become stable only once the region in which the function is approximately linear is included. Thus, to achieve reasonable convergence and accurate measurements of $\delta$ and $\frac{\langle B_t^2\rangle}{\langle B_o^2 \rangle}$ independent of fitting choices, one must capture the approximate plateau or large-scale regime in the structure function. If such a regime is not captured in the structure function due to a lack of data points, the fitting parameters may be inaccurate representations of the true underlying B-field morphology.

Another method, as done in \citet{Hou09}, is to select a value of $l$ and treat the structure function above this as solely a large-scale component. This large-scale component may be fitted separately without the Gaussian term in equation \ref{eq::struct}). The difference between this fitted model and the data can subsequently be fitted using only the Gaussian term of equation \ref{eq::struct}. As the choice of $l$ used to separate the large-scale and small-scale is arbitrary, we investigate the effect of this choice as done above for the maximum $l$ using the $\alpha=-2$ and $k_{\rm{min}}=1$ maps. We find the same general trend that the estimation of $\delta$ is dependent on this separation $l$ until $l\sim30$; however the ratio $\frac{\langle B_t^2\rangle}{\langle B_o^2 \rangle}$ is less impacted, likely due to it acting as the dominant scaling term if $\delta>W$, as is the case here, and so is strongly constrained purely by the maximum value of $1 - \langle\cos{\Delta\psi(l)}\rangle$.

To investigate the effect of pixel size on the fitting of the structure function, we fit all 50 random realisations of the $k_{\rm{min}}=1$ maps with $\alpha=-1,-2,-4$ out to a maximum of $l=100$ and present the cumulative distributions of the fitting parameters $\delta$ and $\frac{\langle B_t^2\rangle}{\langle B_o^2 \rangle}$ in figure \ref{fig::struct_pixel}. There is a considerable reduction in the measured value of $\frac{\langle B_t^2\rangle}{\langle B_o^2 \rangle}$ when increasing pixel size for morphologies with considerable dispersion on smaller scales (i.e. $\alpha=-1$), as expected by the reduction in amplitude of the structure function in figure \ref{fig::struct}.

Concerning the estimation of $\delta$, one sees that dispersion maps with more negative values of $\alpha$ have larger values of $\delta$; as expected as such maps have a greater fraction of dispersion on larger length-scales. Further, for all cases of $\alpha$, the estimated value of $\delta$ increases with increasing pixel size as information on smaller scales is lost, with this biasing most apparent for morphologies with smaller scale dispersion. Thus, the pixel size, B-field morphology, and the fitting method will all affect estimates of the magnetic field strength, and consequently the importance of the magnetic field in the dynamics of a given structure.

It should also be noted that the fitting equation (equation \ref{eq::struct}), cannot adequately fit the more complex oscillatory pattern seen when an maximum scale of dispersion is present ($k_{\rm{min}}>1$). Nor, as can be seen in figure \ref{fig::struct_fit}, when the fit has converged for the $k_{\rm{min}}=1$ cases, it does not fully capture the turnover regime (around $l\sim15-30$ in this example); the fit is overly sharp compared to the smoother transition between the steep and shallow regimes of the structure function. 

\section{Discussion}\label{SEC:DIS}%
From the results presented above, and from prior works \citep[e.g.][]{Hei01,WieWat04,Fal08,Hou09,PatFis19} it is apparent that larger beams and pixels always lead to the an underestimation of the angular dispersion of a given B-field morphology, regardless of if an unsharp-masking method or structure function approach is used. 

Here we have focused on how the degree of this effect is dependent on the exact nature of the dispersion, with cases where a significant dispersion is found at small scales being most sensitive. The use of varying filter sizes in the unsharp-masking method and the shape of the angular dispersion structure function, can both be used to determine the balance between dispersion of larger scales versus smaller scales in a qualitative sense, and thus provide a method to gauge underestimation of the angular dispersion via the correction factor introduced in section \ref{SSEC:FILTER}. Further, while the beam size is fixed by the observations, we do find that the use of sub-beam pixel sizes lessens the underestimation of the angular dispersion in all cases, and thus, where possible, recommend their use.

The most important consequence of this consistent underestimation of the angular dispersion is on the determination of the magnetic-field strength using the Davis-Chandrasekhar-Fermi method \citep{Dav51,ChaFer53} and its variants \citep{Hil09,Hou09,Ska21}. The equation for the plane-of-sky magnetic field strength using the DCF method is:
\begin{equation}
B_{\rm{pos}} \propto \frac{1}{\sigma_\psi},
\end{equation}
and for the structure function method is:
\begin{equation}
B_{\rm{pos}} \propto \left[ \frac{\langle B_t^2\rangle}{\langle B_o^2 \rangle} \right]^{-1/2}.
\end{equation}
Thus, the consistent underestimation of the angular dispersion or the ratio $\frac{\langle B_t^2\rangle}{\langle B_o^2 \rangle}$, leads to a chronic overestimation of the magnetic field strength. This will have a consequence on one's understanding of the role of the B-field in the energetics of star-forming regions, erroneously skewing our conceptualisation of the star formation process to a more magnetically dominated one.

Interestingly, \citet{Pat23} have recently shown that the DCF measurements of magnetic field strengths at a given density are on average approximately 3-5 times larger than those derived from Zeeman splitting measurements. The correction factors found above, suggest that the angular dispersion is typically underestimated by factors of 1-10 when measured on the scale of 1-3x the beam size - commonly taken filter sizes in observations. It is possible that this consistent underestimation of the true angular dispersion used for the DCF method may be a contributing cause to this difference. To combat this, filter-size plots allow one to estimate a correction factor; though this method makes the assumption of a single power-law describing the polarisation angle field both above and below the beam size. Without this correction factor, B-field strength estimates obtained using the classical DCF or structure function methods should be considered overestimates and thus treated as upper limits rather than faithful representations.\footnote{This discussion makes the implicit assumption that the DCF method and its variants are applicable to a given observational dataset. Though this is commonly assumed, this may not be the case. See \citet{Liu22} and \citet{Pat23} for more details.}

\begin{figure*}
\centering
\includegraphics[width=0.9\linewidth,trim={0cm 0cm 20cm 0cm},clip]{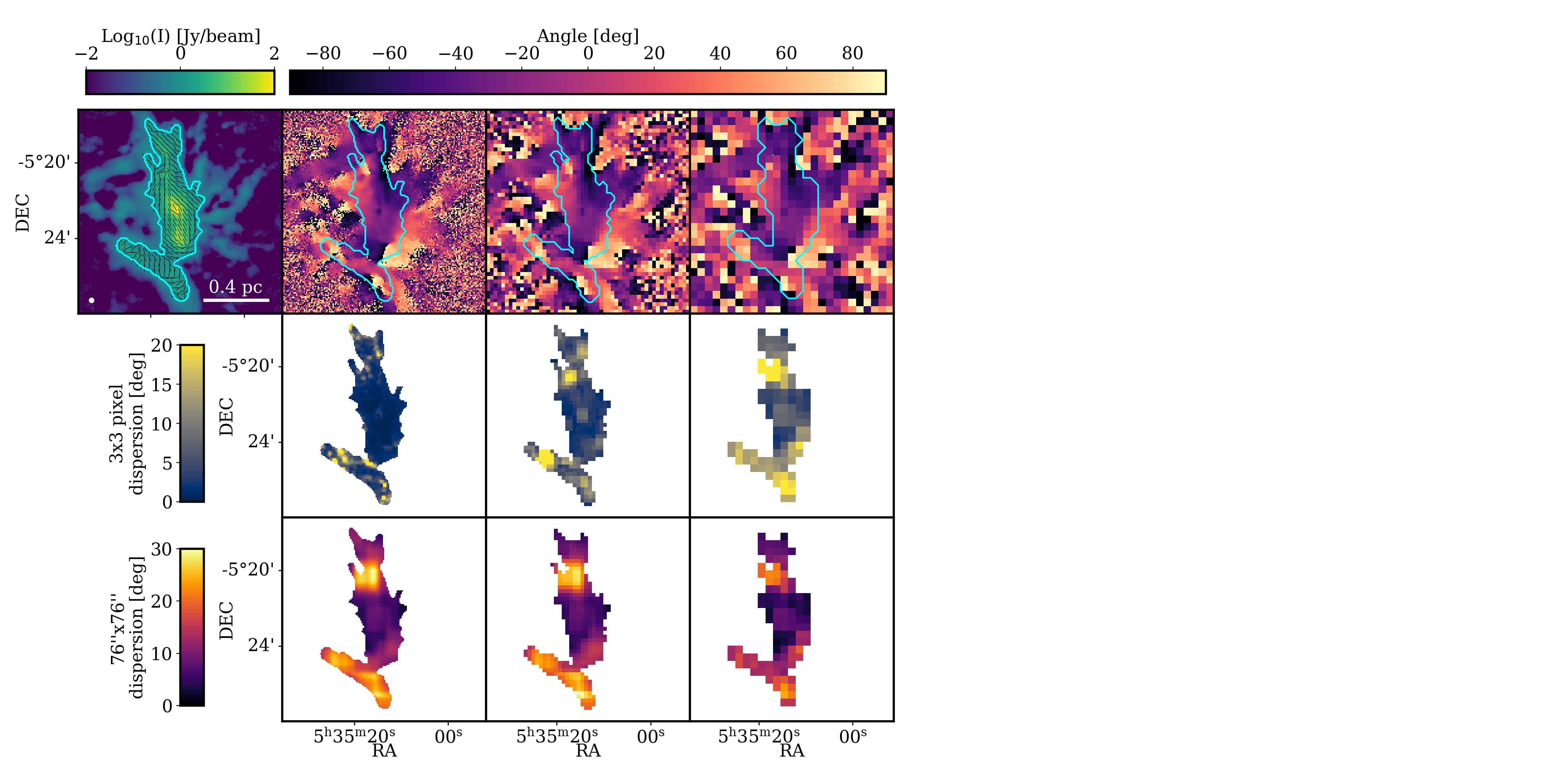}
\caption{(Top left panel) The 850 $\mu$m Stokes I map with a contour showing the 0.7 Jy/beam enclosed region considered in the analysis and black segments showing the polarisation segments. The white circle in the lower left corner shows the 14'' beam size for the JCMT at 850 $\mu$m. (Top row) The polarisation segment position angle determined at (left column) 4'', (middle column) 12'', (right column) 24'' pixel sizes. (Middle row) The local angular dispersion determined using a 3x3 pixel filter size for the unsharp-masking method. (Bottom row) The local angular dispersion determined using an approximately same size filter size for each pixel size: 76'', 84'' and 72'' respectively.}
\label{fig::orion_map}
\end{figure*}  

Additionally, as shown in the above section, the scale at which the angular dispersion is measured, either the filter size in the unsharp-masking method or the maximum extent covered by the structure function fitting, has a profound impact on the angular dispersion estimate and so B-field strength determination. When comparing between different star-forming regions, each being observed at different distances and so physical resolutions, the use of a, for example, 3x3 pixel filter size for the unsharp-masking method leads to considerable difficulty as it corresponds to differing physical scales. Differences between the measured angular dispersion, and thus B-field strengths, of these regions may then be primarily due to differences in scales rather than underlying B-field morphology. It would be beneficial to thus report the physical scale at which angular dispersion is measured to improve the comparison between regions.

\section{Application to real observations - Orion A}\label{SEC:ORION}%
To emphasise the impact of the effects discussed in section \ref{SEC:RES} on observations, we apply techniques of measuring the angular dispersion to JCMT observations to the Orion A OMC-1 region. The data was taken as part of the JCMT BISTRO survey \citep{War17} and was presented in \citet{Hwa21}. This particular region is selected due to its close proximity, brightness and size allowing for a wide range of pixels to be probed at high signal-to-noise.

Figure \ref{fig::orion_map} shows the JCMT 850 $\mu$m Stokes I intensity map of the Orion A OMC-1, with the region of interest enclosed by a 0.7 Jy/beam contour. The exact value of this boundary is arbitrary but results in a contiguous cloud region which matches the dense OMC-1 region previously studied by others \citep[e.g.][]{Hac17, Bri20, Hwa21}. This map is gridded using 4'' pixels and has a beam size of $\sim$14''. Also included in figure \ref{fig::orion_map} is the polarisation angle maps, $\psi$, derived from the Stokes Q and U observations, at 4'', 12'' and 24'' pixel sizes, with the corresponding $I>0.7$ Jy/beam contours. Considering a distance of 414 pc to Orion A \citep{Men07}, these pixel sizes correspond to physical sizes of 0.008, 0.024 and 0.048 pc respectively. 

\begin{figure}
\centering
\includegraphics[width=0.95\linewidth,trim={0.5cm 0.5cm 0.5cm 0.5cm},clip]{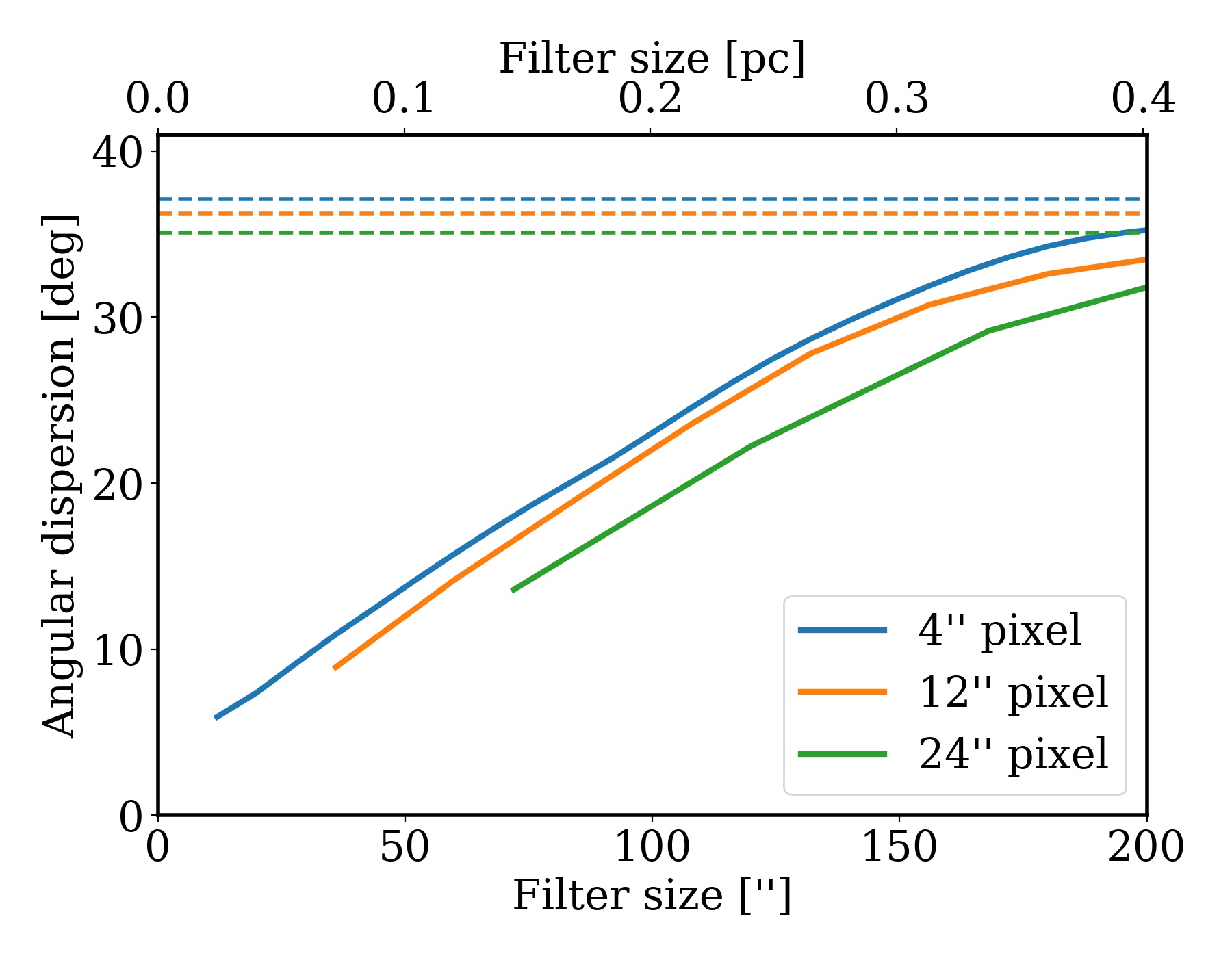}
\caption{The angular dispersion within the intensity contour seen in figure \ref{fig::orion_map} using the unsharp-masking technique with varying filter sizes. The colours denote the pixel sizes used, and the dashed horizontal lines the total angular dispersion across the whole map.}
\label{fig::orion_filter}
\end{figure}

To investigate the effect of the pixel size on these observations, we first use the unsharp-masking method with the commonly selected 3x3 pixel filter size. Considering the angular dispersion of the entire map after using these filters, we find angular dispersions of 5.9$\degree$, 8.9$\degree$ and 13.6$\degree$ respectively. Thus, a change of pixel size from $\sim$1/3 to $\sim$2 times the beam size results in a $>$2 times increase in the measured angular dispersion using a 3x3 pixel filter, and consequently a factor of $\sim$5 decrease in the magnetic field energy ($E_B \propto B^2 \propto 1/\sigma_{\psi}^2$). This increase in the measured angular dispersion with increasing pixel size appears to contradict the assertion made in the prior sections that one expects to see a decrease in measured angular dispersion with increasing pixel size; however, this increase is due to the concurrent increase in filter size - 12'', 36'' and 72'' for the 4'', 12'' and 24'' pixels. This can easily be seen in figure \ref{fig::orion_filter}, which shows the filter-size plot for this region for the three pixel sizes. When dispersion is measured using the same sized filter, there is a noticeable reduction when using larger pixel sizes, as expected from our results above with parametrised maps. For a filter size of 76'' (0.15 pc), the measured angular dispersion is 18.8$\degree$, 17.3$\degree$ and 14.3$\degree$ for the 4'', 12'' and 24'' pixels respectively.\footnote{The values for the 12'' and 24'' pixel maps are derived using linear interpolation as no filter of exactly 76'' can be constructed.}

The examination of the dispersion as a function of pixel size and filter size may be extended by considering the local dispersion rather than the global dispersion \citep[as done in e.g.][]{Hwa21, Ngoc23}. This is done by using the unsharp-masking method but assigning the local dispersion within the filter to the central pixel rather than calculating a residual angle. Note that due to the possibility of a low number of angles within a filter, a common issue of observations with low signal-to-noise, which is known to lead to underestimates of dispersion, we use the recommended factor of 1/$\sqrt{N-1.5}$ instead of 1/$\sqrt{N-1}$ for the standard deviation calculation \citep{Gur71}\footnote{This statistical factor is an approximation to a more complete factor but is accurate to $\lesssim3\%$ for $N>3$.}. 

The results for 3x3 pixel and approximately 76''x76'' filter sizes can be seen in the middle and bottom row respectively of figure \ref{fig::orion_map}. One sees that in OMC-1, levels of high angular dispersion are predominately located in the Orion bar region, returning values of $\sim10-20$ degrees compared to the $\sim3-6$ degrees typical towards the central region when measured using 4'' pixels and a 12'' filter size. As with the globally calculated dispersions reported above, the local dispersion sees a decrease with increasing pixel size at approximately the same filter size regardless of the region of the map.

\begin{figure}
\centering
\includegraphics[width=0.95\linewidth,trim={0.5cm 0.5cm 0.5cm 0.5cm},clip]{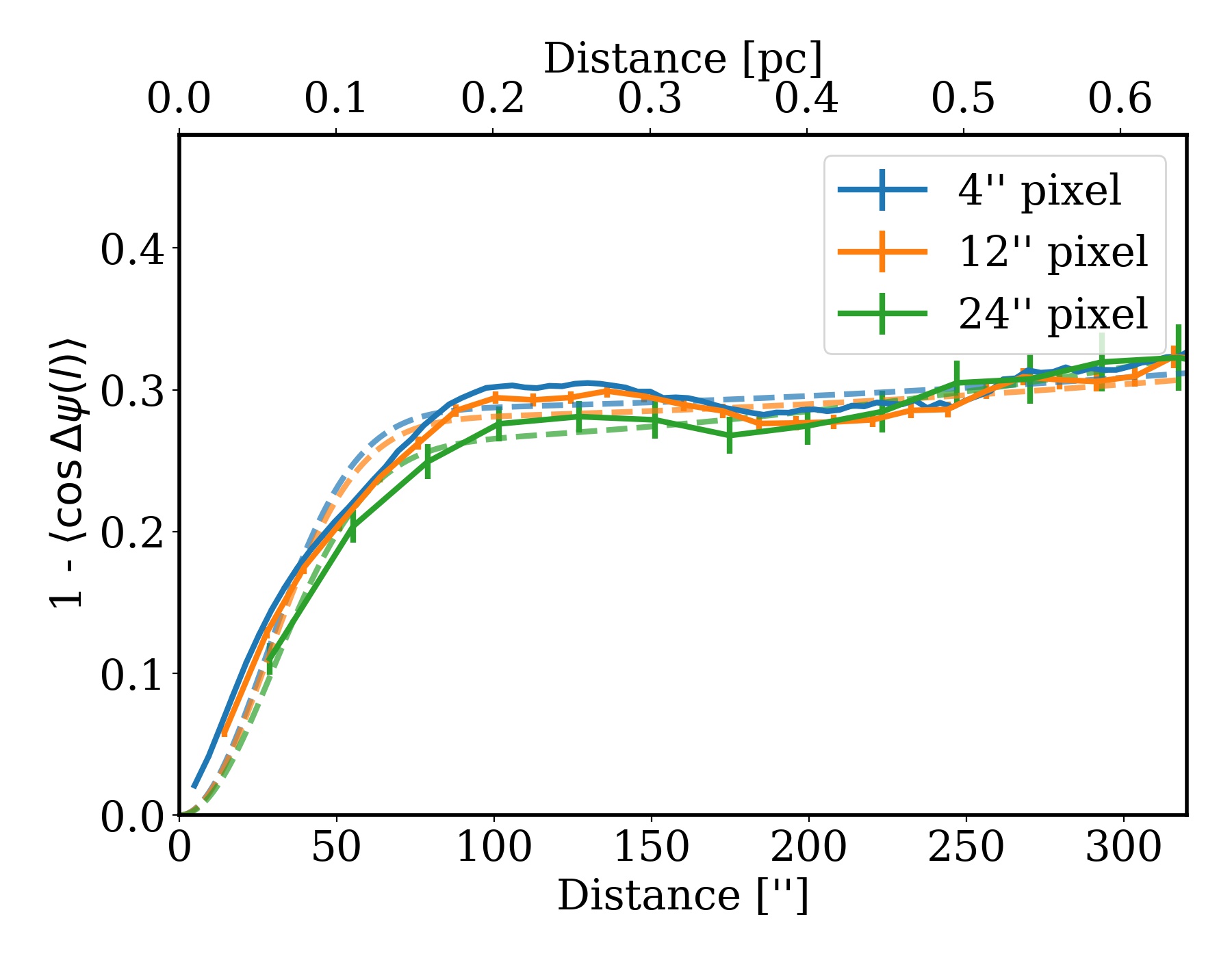}
\caption{The structure function calculated using the region enclosed by the intensity contour seen in figure \ref{fig::orion_map} using 4'', 12'' and 24'' pixel maps (blue, orange and green respectively). The dashed lines of the same colour show the fit using equation \ref{eq::struct}. The structure function has been binned using pixel-sized bins, and the mean distance within that bin has been plotted. The error bars show the error on the mean.}
\label{fig::orion_struct}
\end{figure}  

\begin{figure*}
\centering
\includegraphics[width=0.9\linewidth,trim={2.5cm 0cm 4.7cm 1cm},clip]{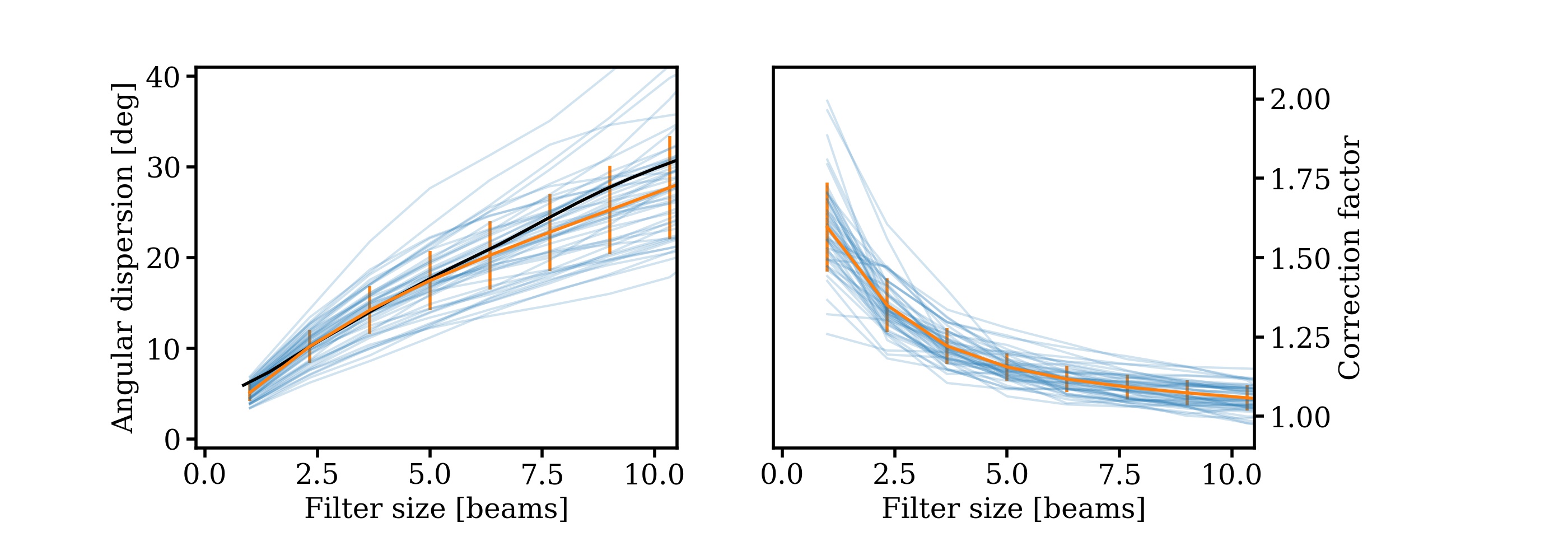}
\caption{(Left) The angular dispersion measured as a function of the filter size in terms of beam size for (blue lines) 50 random realisations of an angle map characterised by $\alpha=-3$, a beam size of 3, total dispersion of 60$\degree$, and a grid size of 60. The mean and standard deviation of this set is shown in orange, with the JCMT OMC-1 results shown in black. (Right) The correction factor as a function of filter size for the same angle maps as the left panel.}
\label{fig::orion_correction}
\end{figure*}  

To further investigate the multi-scale dispersion in the map of OMC-1, we show the structure function calculated for the three pixel sizes in figure \ref{fig::orion_struct}. The shape of the three functions are similar, with only a slight decrease in their magnitude with increasing pixel sizes, as also seen in the parametrised maps of section \ref{SSEC:STRUCT}. Moreover, obvious oscillations do not appear in the structure function, suggesting the lack of a maximum/dominant dispersion scale, as seen in the $k_{\rm{min}}>1$ cases, indicating that dispersion continues to increase with scale up to the map size.  Fitting the structure function out to a distance of 300'' using equation \ref{eq::struct} returns values of $\delta$ of $25.8\pm0.8$'', $26.2\pm1.3$'', and $28.0\pm1.5$'' for the 4'', 12'', and 24'' pixel maps respectively, and values of $\frac{\langle B_t^2\rangle}{\langle B_o^2 \rangle}$ of $1.27\pm0.04$, $1.21\pm0.07$, and $1.04\pm0.06$\footnote{Note that a value of $L$, the thickness of the structure along the line-of-sight, must be assumed for equation $4$ to be fitted. Here a value of 250'' is used, which is approximately the equivalent radius of the region enclosed by the 0.7 Jy/beam contour. The fitting parameter $\frac{\langle B_t^2\rangle}{\langle B_o^2 \rangle}$ is linearly related to the assumed value of $L$, i.e. a reduction in $L$ by a factor of 2 would reduce $\frac{\langle B_t^2\rangle}{\langle B_o^2 \rangle}$ by a factor of 2.}. This slight increase in $\delta$ and decrease in $\frac{\langle B_t^2\rangle}{\langle B_o^2 \rangle}$ with increasing pixel size is also expected from our study of parametrised maps. 

The filter-size plot in figure \ref{fig::orion_filter} and the structure function in figure \ref{fig::orion_struct}, and the moderate reduction in angular dispersion with increasing pixel size measured with a uniform filter size and in the fitted parameter $\frac{\langle B_t^2\rangle}{\langle B_o^2 \rangle}$ with increasing pixel size, all suggest that the OMC-1 region is similar to the $\alpha=-2$ case. Such a picture is consistent with what can be seen visually in figure \ref{fig::orion_map}, where both moderate and large scale variations in the polarisation angle $\psi$ are apparent.

\subsection{Correction factor}\label{SSEC:CORRECT}%

To estimate a correction factor to determine how significantly the angular dispersion is being underestimated in the OMC-1 region using the unsharp-masking method, we first calculate a set of parametrised maps which approximate the filter-size plot derived from the JCMT data (figure \ref{fig::orion_filter}). This is done by varying $\alpha$ between -2 and -4 in steps of 0.5, the beam size from 3 to 12 in steps of 3, the total dispersion from 40 to 60 degrees in steps of 10 degrees, and the grid size (i.e. the largest length-scale on which dispersion occurs) from 60 to 120 in steps of 30. The result is that a map of $\alpha=-3$, a beam size of 3, total dispersion of 60$\degree$, and a grid size of 60, best approximates the OMC-1 data. The left panel of figure \ref{fig::orion_correction} shows the result of 50 random realisations of such a polarisation angle map, and their mean. We do not assert here that the B-field morphology in OMC-1 is governed by exactly such a power-law, but that fields generated with these parameters adequately describe the angular dispersion in OMC-1's B-field across multiple scales and so provide a manner to estimate a correction factor.

The right panel of figure \ref{fig::orion_correction} shows the correction factor for the 50 random realisations, as well as its mean and standard deviation. At a filter size equivalent to one beam (14''), the correction factor is 1.60 $\pm$ 0.14; for a filter size equal to three beams (42''), the correction factor is 1.29 $\pm$ 0.07. The particularly low correction factor is the result of two properties of OMC-1's B-field morphology. First, the angle map is well described by a steep power-law $\alpha=-3$ and so considerable dispersion is found on scales larger than the beam and thus little affected by beam convolution (median dispersion scale $\sim$10 beam sizes). This is constrained by the near linear relationship between filter size and measured angular dispersion in figure \ref{fig::orion_filter} without signs of a plateau until filter sizes of $>10$ beams are reached. Second, that the beam size of the observations is not considerably larger than the smallest scale on which dispersion is to be found, i.e. there is not a significant level of angular dispersion found at scales much smaller than the beam. This is constrained by the relatively large angular dispersion measured when the filter size is equal to one beam size while simultaneously having a steep power-law governing the entire map. Importantly these correction factors mean that using the DCF method, there would be a factor of 1.60 and 1.29 reduction in the B-field strengths measured at those scales, or a factor of 2.56 and 1.66 reduction in the magnetic field energy, compared to the uncorrected values. 

Considering these correction factors, the 4'' pixel size dispersion measured at 14'' and 42'' filter sizes should be converted from 6.3$\degree$ and 12.1$\degree$ to 10.0$\degree$ and 15.6$\degree$ respectively. This prediction could be tested by higher resolution polarisation observations (beam size $\sim 1-2''$) of this region which adequately capture scales greater than these filter sizes. If angular dispersion at this higher resolution is found to be similar to the uncorrected values from the JCMT observations, it would suggest there is a break in the multi-scale balance of dispersion at scales $<14$'', i.e. $\alpha=-3$ no longer holds as it seems to for scales $>14$''.   

\section{Conclusions}\label{SEC:CON}%

The wealth of polarised dust emission observations in recent years has led to a greater understanding of the role of magnetic fields in the interstellar medium; however, the measurement of the angular dispersion, necessary for magnetic field strength estimates when using the DCF method and its variants, is non-trivial for complex field structures \citep[as shown in prior works][]{WieWat04,Hou09,Hou13,PatFis19}. In this paper we present a systematic review of how observational effects (pixel sizes and beam convolution) affect the measurement of angular dispersion in parametrised angle maps with multi-scale dispersion using the commonly employed unsharp-masking and structure function methods.

In agreement with previous works, we show that beam convolution and larger pixel sizes both lead to an underestimate in the measured angular dispersion, with the degree to which it is underestimated dependent on the relative power of dispersion at large and small spatial scales (characterised here by the power-law index $\alpha$ and $k_{\rm{min}}$, sections \ref{SSEC:BEAM} and \ref{SSEC:ALPHA}). When a greater share of dispersion is found close to the beam or pixel size, the measurement of angular dispersion is more greatly reduced. As such, measured angular dispersion should be treated as lower limits. We find that selecting sub-beam sized pixels leads to a measured angular dispersion closer to the true value before beam convolution, and so is encouraged for observations if signal-to-noise allows.

As angular dispersion can be found on multiple size scales, the length-scale at which the dispersion is measured strongly affects the result. Two commonly employed techniques which focus on extracting an angular dispersion on small-scales, i.e. to be used for DCF estimates of magnetic field strengths, are the unsharp-masking method and the structure function. Both methods lead to underestimates due to beam convolution, and increasing pixel size increases the underestimation (section \ref{SSEC:FILTER} and \ref{SSEC:STRUCT}). For the unsharp-masking method, as the measured angular dispersion is heavily dependent on the filter size used, we encourage that the physical length-scale associated with the filter should be reported with the angular dispersion in acknowledgement of this. For the structure function method, the fitting parameters ($\delta$ and $\frac{\langle B_t^2\rangle}{\langle B_o^2 \rangle}$) only converge once the approximate plateau or large-scale regime is included in the fit. It is thus necessary to adequately recover this larger spatial scale for structure function fitting to be used to estimate magnetic field strengths.

Using parametrised angle map, we show that the typical degree of underestimation of the angular dispersion using the unsharp-masking method is a factor of 1-10 when measured at 1-3x beam size. This will have a subsequent impact of overestimating the magnetic field strength using the DCF method by the same factor, leading to a potentially overly magnetically-dominated picture of a region. Interestingly, \citet{Pat23}  show that DCF measurements of magnetic field strengths at a given density are on average approximately 3-5 times larger than those derived from Zeeman splitting measurements. The chronic underestimation of the angular dispersion when measured at scales similar to the beam size, as is commonly done, may be one reason for the discrepancy in estimated magnetic field strengths from these two methods.

By using the information encoded in the filter-size plots (measured angular dispersion against filter size for the unsharp-masking method) we develop a method to estimate a `correction factor' to bring the measured angular dispersion into better alignment with the true value prior to beam convolution. We apply this method to JCMT polarisation data of the Orion A OMC-1 region \citep{Hwa21}, and show that the region's magnetic field structure is well represented by a field with dispersion predominately on scales much larger than the beam size, and with the smallest turbulent length-scale only a factor of 2-3 smaller than the beam size (section \ref{SEC:ORION}). The leads to a relatively small correction factor of only 1.60 and 1.29 when dispersion is measured on 14'' and 42'' scales (0.028 and 0.084 pc respectively) for this region. Due to these correction factors, we predict that higher-resolution polarisation observations on this region ought to find angular dispersions of approximately 10.0$\degree$ and 15.6$\degree$ at 14'' and 42'' scales respectively.

To aid the community, we provide an open-source \textsc{Python} package to produce the parametrised turbulent angle maps shown here, calculate and plot filter-size plots, and estimate the correction factor. The version used in this manuscript can be found at \citet{MCF_soft}, with on-going development at https://github.com/SeamusClarke/MCF

\begin{acknowledgments}
The authors thank Jihye Hwang for providing the JCMT OMC-1 polarization data. SDC and YWT are supported by National Science and Technology Council (NSTC) grants 112-2112-M-001-066 and 111-2112-M-001-064. PMK acknowledges support from NSTC grants NSTC 113-2112-M-001-016-, NSTC 112-2112-M-001-049-, and NSTC 111-2112-M-001-070-. GAF acknowledges support from the Collaborative Research Centre 956, funded by the Deutsche Forschungsgemeinschaft (DFG) project ID 184018867. SDC would also like to acknowledge the people developing and maintaining the open source packages which were used in this work: \textsc{Matplotlib} \citep{matplotlib}, \textsc{NumPy} \citep{numpy}, \textsc{Astropy} \citep{astropyI,astropyII,astropyIII} and \textsc{SciPy} \citep{scipy}.
\end{acknowledgments}

%

\vspace{5mm}
\facilities{JCMT}



\appendix

\section{Periodic, turbulent-box MHD simulations without gravity}\label{SEC:APP}%
To test the assertion made in section \ref{SEC:METHOD} that MHD turbulence may impart a multi-scale dispersion upon the polarisation/B-field angle field which may be characterised as quasi-power-law in Fourier space, we run ideal-MHD simulations using the code \textsc{GAMER-2} \citep{Sch18}. We use a periodic cube setup with a side length of 1 pc and a uniform grid with 256$^3$ cells, resulting in a resolution of $\Delta x \sim 0.004$ pc. Constrained transport is utilised by \textsc{GAMER-2} to ensure a divergence-free magnetic field everywhere. Gravity is not included so that only the effects of ideal-MHD turbulence may be considered; it's inclusion may lead to differences in magnetic-field structure in strongly self-gravitating regions.

The initial density of the setup is 100 cm$^{-3}$ and an isothermal equation state is used with the gas temperature set to 10 K; with a mean particle mass of 2.35, this leads to a sound speed of $\sim$0.19 km/s. A turbulent field is generated using the procedure described in \citet{Lom15}, with a exponent of $\alpha=-4$ and modes between $k=1-128$ populated. The initial Mach number of the turbulence is set to 5, i.e. a velocity dispersion of $\sim 0.94$ km/s. The turbulence is not driven and so is allowed to decay.

The initial magnetic field is set to (0,0,B$_z$) everywhere, where B$_z$ is set to 1 $\mu$G or 4 $\mu$G. At the initial density, this equates to Alfv\'en speeds of $\sim 0.5$ km/s and $\sim 2.0$ km/s respectively. Considering the turbulent velocity, this leads to Alfv\'en Mach numbers of $\sim 1.9$ and $\sim 0.5$, i.e. super-Alfv\'enic and sub-Alfv\'enic turbulence respectively.

\begin{figure*}
\centering
\includegraphics[width=0.99\linewidth,trim={1cm 0.0cm 5cm 0.2cm},clip]{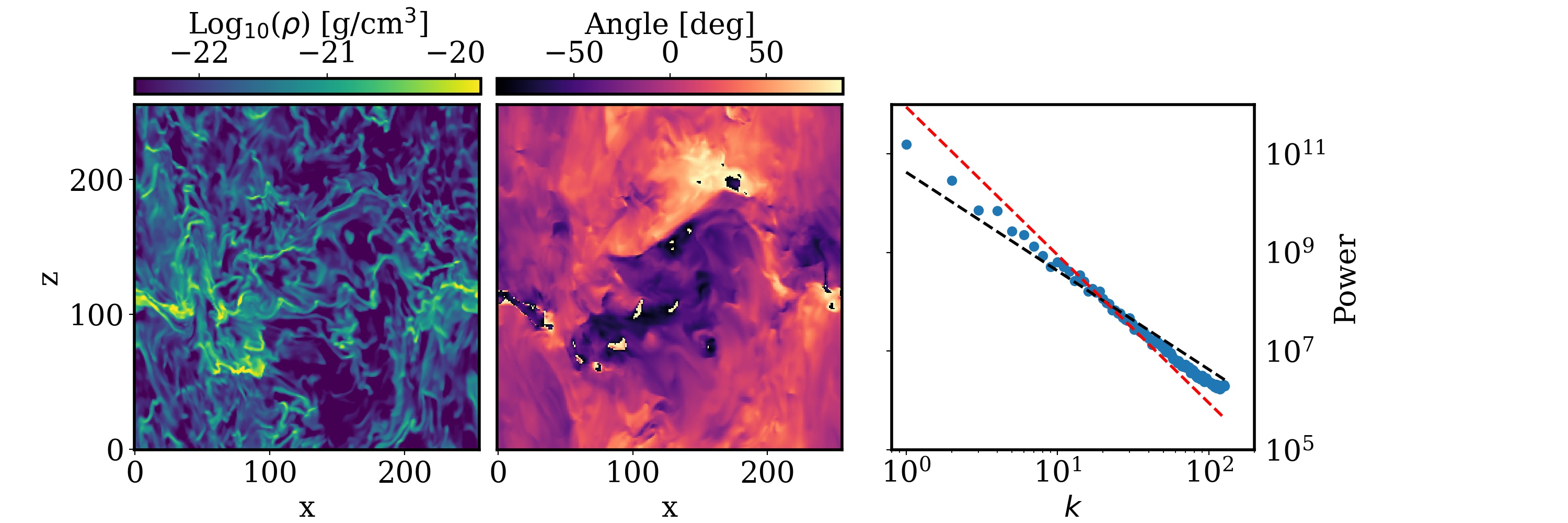}
\includegraphics[width=0.99\linewidth,trim={1cm 0.0cm 5cm 0.2cm},clip]{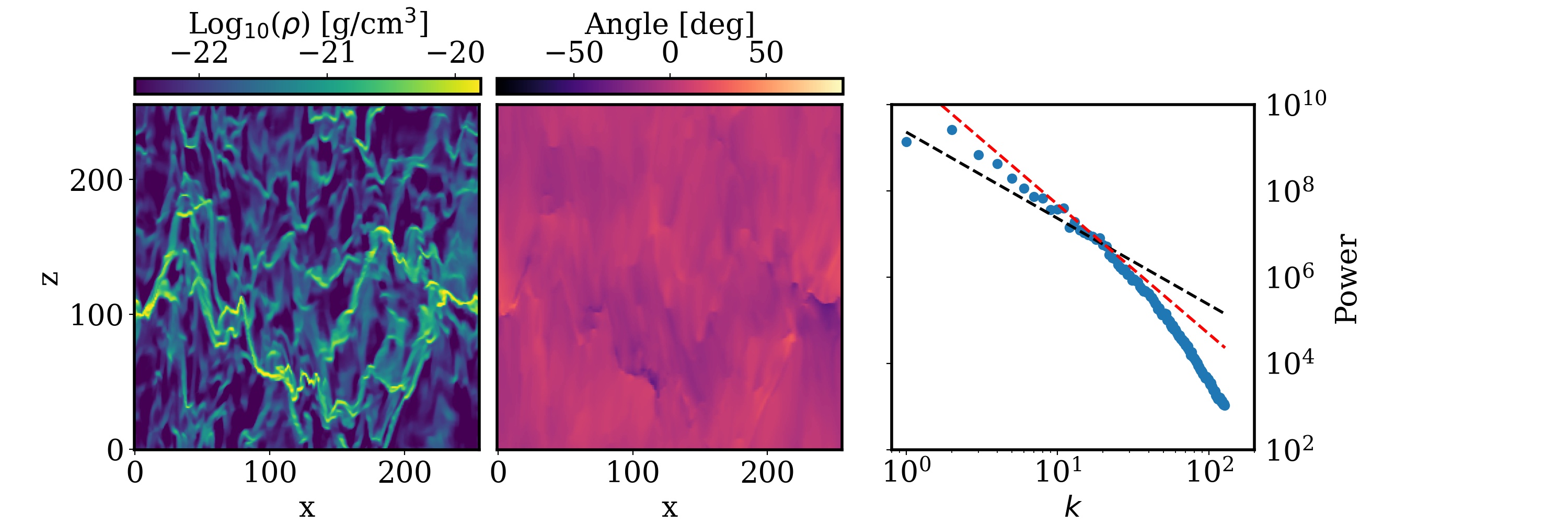}
\caption{Slice plots at $y=0.5$ pc of the (top) super-Alfv\'enic and (bottom) sub-Alfv\'enic simulations at $t=0.5$ Myr. Panels show (left) volume density, (centre) B-field vector angle, and (right) power spectrum resulting from the Fourier transform of the angle map. The black and red dashed lines in the right panels show power-laws with indices of -2 and -3 respectively.}
\label{fig::appendix_MHD}
\end{figure*}

We select the $y=0.5$ pc plane of the cube to investigate the magnetic-field angle, although the results are independent of this choice. Figure \ref{fig::appendix_MHD} shows both simulations at $t=0.5$ Myr $\sim 0.5$ $t_{tc}$, where $t_{tc}$ is the turbulent crossing time. In both the super- and sub-Alfv\'enic the density field has been strongly perturbed and produced a filamentary medium, with corresponding perturbations in the angle map of the B-field - angles are calculated using the $x$ and $z$ components of the B-field with respect to the $z$-axis and range from -90 to 90 degrees. The power spectra resulting from the Fourier transform of these angle maps can be seen in the rightmost panels. In the super-Alfv\'enic simulation, a single power-law is apparent across all spatial scales with a power-law index of between -2 and -3. For the sub-Alfv\'enic simulation, the clear anisotropy introduced by the stronger magnetic field leads to two power-laws. These results are also found if later time points are considered, though as the turbulence decays the power-law steepens and the total amplitude reduces. 

Taken together, it is clear that in both the super- and sub-Alfv\'enic cases, the complex magnetic field morphology produced by MHD turbulence can be characterised by a power-law like model across a wide range of spatial scales, supporting the use of such models in the main body of this manuscript.

\bibliography{ref}{}

\begin{thebibliography}{}
\expandafter\ifx\csname natexlab\endcsname\relax\def\natexlab#1{#1}\fi
\providecommand{\url}[1]{\href{#1}{#1}}
\providecommand{\dodoi}[1]{doi:~\href{http://doi.org/#1}{\nolinkurl{#1}}}
\providecommand{\doeprint}[1]{\href{http://ascl.net/#1}{\nolinkurl{http://ascl.net/#1}}}
\providecommand{\doarXiv}[1]{\href{https://arxiv.org/abs/#1}{\nolinkurl{https://arxiv.org/abs/#1}}}

\bibitem[{{Andersson} {et~al.}(2015){Andersson}, {Lazarian}, \&
  {Vaillancourt}}]{And15}
{Andersson}, B.~G., {Lazarian}, A., \& {Vaillancourt}, J.~E. 2015, \araa, 53,
  501, \dodoi{10.1146/annurev-astro-082214-122414}

\bibitem[{{Astropy Collaboration} {et~al.}(2013){Astropy Collaboration},
  {Robitaille}, {Tollerud}, {Greenfield}, {Droettboom}, {Bray}, {Aldcroft},
  {Davis}, {Ginsburg}, {Price-Whelan}, {Kerzendorf}, {Conley}, {Crighton},
  {Barbary}, {Muna}, {Ferguson}, {Grollier}, {Parikh}, {Nair}, {Unther},
  {Deil}, {Woillez}, {Conseil}, {Kramer}, {Turner}, {Singer}, {Fox}, {Weaver},
  {Zabalza}, {Edwards}, {Azalee Bostroem}, {Burke}, {Casey}, {Crawford},
  {Dencheva}, {Ely}, {Jenness}, {Labrie}, {Lim}, {Pierfederici}, {Pontzen},
  {Ptak}, {Refsdal}, {Servillat}, \& {Streicher}}]{astropyI}
{Astropy Collaboration}, {Robitaille}, T.~P., {Tollerud}, E.~J., {et~al.} 2013,
  \aap, 558, A33, \dodoi{10.1051/0004-6361/201322068}

\bibitem[{{Astropy Collaboration} {et~al.}(2018){Astropy Collaboration},
  {Price-Whelan}, {Sip{\H{o}}cz}, {G{\"u}nther}, {Lim}, {Crawford}, {Conseil},
  {Shupe}, {Craig}, {Dencheva}, {Ginsburg}, {VanderPlas}, {Bradley},
  {P{\'e}rez-Su{\'a}rez}, {de Val-Borro}, {Aldcroft}, {Cruz}, {Robitaille},
  {Tollerud}, {Ardelean}, {Babej}, {Bach}, {Bachetti}, {Bakanov}, {Bamford},
  {Barentsen}, {Barmby}, {Baumbach}, {Berry}, {Biscani}, {Boquien}, {Bostroem},
  {Bouma}, {Brammer}, {Bray}, {Breytenbach}, {Buddelmeijer}, {Burke},
  {Calderone}, {Cano Rodr{\'\i}guez}, {Cara}, {Cardoso}, {Cheedella}, {Copin},
  {Corrales}, {Crichton}, {D'Avella}, {Deil}, {Depagne}, {Dietrich}, {Donath},
  {Droettboom}, {Earl}, {Erben}, {Fabbro}, {Ferreira}, {Finethy}, {Fox},
  {Garrison}, {Gibbons}, {Goldstein}, {Gommers}, {Greco}, {Greenfield},
  {Groener}, {Grollier}, {Hagen}, {Hirst}, {Homeier}, {Horton}, {Hosseinzadeh},
  {Hu}, {Hunkeler}, {Ivezi{\'c}}, {Jain}, {Jenness}, {Kanarek}, {Kendrew},
  {Kern}, {Kerzendorf}, {Khvalko}, {King}, {Kirkby}, {Kulkarni}, {Kumar},
  {Lee}, {Lenz}, {Littlefair}, {Ma}, {Macleod}, {Mastropietro}, {McCully},
  {Montagnac}, {Morris}, {Mueller}, {Mumford}, {Muna}, {Murphy}, {Nelson},
  {Nguyen}, {Ninan}, {N{\"o}the}, {Ogaz}, {Oh}, {Parejko}, {Parley}, {Pascual},
  {Patil}, {Patil}, {Plunkett}, {Prochaska}, {Rastogi}, {Reddy Janga},
  {Sabater}, {Sakurikar}, {Seifert}, {Sherbert}, {Sherwood-Taylor}, {Shih},
  {Sick}, {Silbiger}, {Singanamalla}, {Singer}, {Sladen}, {Sooley},
  {Sornarajah}, {Streicher}, {Teuben}, {Thomas}, {Tremblay}, {Turner},
  {Terr{\'o}n}, {van Kerkwijk}, {de la Vega}, {Watkins}, {Weaver}, {Whitmore},
  {Woillez}, {Zabalza}, \& {Astropy Contributors}}]{astropyII}
{Astropy Collaboration}, {Price-Whelan}, A.~M., {Sip{\H{o}}cz}, B.~M., {et~al.}
  2018, \aj, 156, 123, \dodoi{10.3847/1538-3881/aabc4f}

\bibitem[{{Astropy Collaboration} {et~al.}(2022){Astropy Collaboration},
  {Price-Whelan}, {Lim}, {Earl}, {Starkman}, {Bradley}, {Shupe}, {Patil},
  {Corrales}, {Brasseur}, {N{\"o}the}, {Donath}, {Tollerud}, {Morris},
  {Ginsburg}, {Vaher}, {Weaver}, {Tocknell}, {Jamieson}, {van Kerkwijk},
  {Robitaille}, {Merry}, {Bachetti}, {G{\"u}nther}, {Aldcroft},
  {Alvarado-Montes}, {Archibald}, {B{\'o}di}, {Bapat}, {Barentsen},
  {Baz{\'a}n}, {Biswas}, {Boquien}, {Burke}, {Cara}, {Cara}, {Conroy},
  {Conseil}, {Craig}, {Cross}, {Cruz}, {D'Eugenio}, {Dencheva}, {Devillepoix},
  {Dietrich}, {Eigenbrot}, {Erben}, {Ferreira}, {Foreman-Mackey}, {Fox},
  {Freij}, {Garg}, {Geda}, {Glattly}, {Gondhalekar}, {Gordon}, {Grant},
  {Greenfield}, {Groener}, {Guest}, {Gurovich}, {Handberg}, {Hart},
  {Hatfield-Dodds}, {Homeier}, {Hosseinzadeh}, {Jenness}, {Jones}, {Joseph},
  {Kalmbach}, {Karamehmetoglu}, {Ka{\l}uszy{\'n}ski}, {Kelley}, {Kern},
  {Kerzendorf}, {Koch}, {Kulumani}, {Lee}, {Ly}, {Ma}, {MacBride}, {Maljaars},
  {Muna}, {Murphy}, {Norman}, {O'Steen}, {Oman}, {Pacifici}, {Pascual},
  {Pascual-Granado}, {Patil}, {Perren}, {Pickering}, {Rastogi}, {Roulston},
  {Ryan}, {Rykoff}, {Sabater}, {Sakurikar}, {Salgado}, {Sanghi}, {Saunders},
  {Savchenko}, {Schwardt}, {Seifert-Eckert}, {Shih}, {Jain}, {Shukla}, {Sick},
  {Simpson}, {Singanamalla}, {Singer}, {Singhal}, {Sinha}, {Sip{\H{o}}cz},
  {Spitler}, {Stansby}, {Streicher}, {{\v{S}}umak}, {Swinbank}, {Taranu},
  {Tewary}, {Tremblay}, {de Val-Borro}, {Van Kooten}, {Vasovi{\'c}}, {Verma},
  {de Miranda Cardoso}, {Williams}, {Wilson}, {Winkel}, {Wood-Vasey}, {Xue},
  {Yoachim}, {Zhang}, {Zonca}, \& {Astropy Project Contributors}}]{astropyIII}
{Astropy Collaboration}, {Price-Whelan}, A.~M., {Lim}, P.~L., {et~al.} 2022,
  \apj, 935, 167, \dodoi{10.3847/1538-4357/ac7c74}

\bibitem[{{Brinkmann} {et~al.}(2020){Brinkmann}, {Wyrowski}, {Kauffmann},
  {Colombo}, {Menten}, {Tang}, \& {G{\"u}sten}}]{Bri20}
{Brinkmann}, N., {Wyrowski}, F., {Kauffmann}, J., {et~al.} 2020, \aap, 636,
  A39, \dodoi{10.1051/0004-6361/201936885}

\bibitem[{{Chandrasekhar} \& {Fermi}(1953)}]{ChaFer53}
{Chandrasekhar}, S., \& {Fermi}, E. 1953, \apj, 118, 113,
  \dodoi{10.1086/145731}

\bibitem[{{Chen} \& {Ostriker}(2015)}]{Che15}
{Chen}, C.-Y., \& {Ostriker}, E.~C. 2015, \apj, 810, 126,
  \dodoi{10.1088/0004-637X/810/2/126}

\bibitem[{Clarke(2026)}]{MCF_soft}
Clarke, S. 2026, Magnetic angular dispersion Correction Factor (MCF) code,
  Zenodo, \dodoi{10.5281/zenodo.19202295}

\bibitem[{{Cudlip} {et~al.}(1982){Cudlip}, {Furniss}, {King}, \&
  {Jennings}}]{Cud82}
{Cudlip}, W., {Furniss}, I., {King}, K.~J., \& {Jennings}, R.~E. 1982, \mnras,
  200, 1169, \dodoi{10.1093/mnras/200.4.1169}

\bibitem[{{Davis}(1951)}]{Dav51}
{Davis}, L. 1951, Physical Review, 81, 890, \dodoi{10.1103/PhysRev.81.890.2}

\bibitem[{{Davis} \& {Greenstein}(1949)}]{DavGre49}
{Davis}, L., \& {Greenstein}, J.~L. 1949, Physical Review, 75, 1605,
  \dodoi{10.1103/PhysRev.75.1605}

\bibitem[{{Falceta-Gon{\c{c}}alves} {et~al.}(2008){Falceta-Gon{\c{c}}alves},
  {Lazarian}, \& {Kowal}}]{Fal08}
{Falceta-Gon{\c{c}}alves}, D., {Lazarian}, A., \& {Kowal}, G. 2008, \apj, 679,
  537, \dodoi{10.1086/587479}

\bibitem[{{Federrath}(2016)}]{Fed16}
{Federrath}, C. 2016, Journal of Plasma Physics, 82, 535820601,
  \dodoi{10.1017/S0022377816001069}

\bibitem[{{Ganguly} {et~al.}(2024){Ganguly}, {Walch}, {Clarke}, \&
  {Seifried}}]{Gan24}
{Ganguly}, S., {Walch}, S., {Clarke}, S.~D., \& {Seifried}, D. 2024, \mnras,
  528, 3630, \dodoi{10.1093/mnras/stae032}

\bibitem[{{Ganguly} {et~al.}(2023){Ganguly}, {Walch}, {Seifried}, {Clarke}, \&
  {Weis}}]{Gan23}
{Ganguly}, S., {Walch}, S., {Seifried}, D., {Clarke}, S.~D., \& {Weis}, M.
  2023, \mnras, 525, 721, \dodoi{10.1093/mnras/stad2054}

\bibitem[{Gurland \& Tripathi(1971)}]{Gur71}
Gurland, J., \& Tripathi, R.~C. 1971, The American Statistician, 25, 30,
  \dodoi{10.1080/00031305.1971.10477279}

\bibitem[{{Hacar} {et~al.}(2017){Hacar}, {Tafalla}, \& {Alves}}]{Hac17}
{Hacar}, A., {Tafalla}, M., \& {Alves}, J. 2017, ArXiv e-prints.
\newblock \doarXiv{1703.07029}

\bibitem[{Harris {et~al.}(2020)Harris, Millman, van~der Walt, Gommers,
  Virtanen, Cournapeau, Wieser, Taylor, Berg, Smith, Kern, Picus, Hoyer, van
  Kerkwijk, Brett, Haldane, del R{\'{i}}o, Wiebe, Peterson,
  G{\'{e}}rard-Marchant, Sheppard, Reddy, Weckesser, Abbasi, Gohlke, \&
  Oliphant}]{numpy}
Harris, C.~R., Millman, K.~J., van~der Walt, S.~J., {et~al.} 2020, Nature, 585,
  357, \dodoi{10.1038/s41586-020-2649-2}

\bibitem[{{Heitsch} {et~al.}(2001){Heitsch}, {Zweibel}, {Mac Low}, {Li}, \&
  {Norman}}]{Hei01}
{Heitsch}, F., {Zweibel}, E.~G., {Mac Low}, M.-M., {Li}, P., \& {Norman}, M.~L.
  2001, \apj, 561, 800, \dodoi{10.1086/323489}

\bibitem[{{Hildebrand}(1988)}]{Hil88}
{Hildebrand}, R.~H. 1988, \qjras, 29, 327

\bibitem[{{Hildebrand} {et~al.}(1984){Hildebrand}, {Dragovan}, \&
  {Novak}}]{Hil84}
{Hildebrand}, R.~H., {Dragovan}, M., \& {Novak}, G. 1984, \apjl, 284, L51,
  \dodoi{10.1086/184351}

\bibitem[{{Hildebrand} {et~al.}(2009){Hildebrand}, {Kirby}, {Dotson}, {Houde},
  \& {Vaillancourt}}]{Hil09}
{Hildebrand}, R.~H., {Kirby}, L., {Dotson}, J.~L., {Houde}, M., \&
  {Vaillancourt}, J.~E. 2009, \apj, 696, 567,
  \dodoi{10.1088/0004-637X/696/1/567}

\bibitem[{{Hiltner}(1949)}]{Hil49}
{Hiltner}, W.~A. 1949, Science, 109, 165, \dodoi{10.1126/science.109.2825.165}

\bibitem[{{Hoang} {et~al.}(2021){Hoang}, {Tram}, {Lee}, {Diep}, \&
  {Ngoc}}]{Hoa21}
{Hoang}, T., {Tram}, L.~N., {Lee}, H., {Diep}, P.~N., \& {Ngoc}, N.~B. 2021,
  \apj, 908, 218, \dodoi{10.3847/1538-4357/abd54f}

\bibitem[{{Houde} {et~al.}(2013){Houde}, {Fletcher}, {Beck}, {Hildebrand},
  {Vaillancourt}, \& {Stil}}]{Hou13}
{Houde}, M., {Fletcher}, A., {Beck}, R., {et~al.} 2013, \apj, 766, 49,
  \dodoi{10.1088/0004-637X/766/1/49}

\bibitem[{{Houde} {et~al.}(2016){Houde}, {Hull}, {Plambeck}, {Vaillancourt}, \&
  {Hildebrand}}]{Hou16}
{Houde}, M., {Hull}, C. L.~H., {Plambeck}, R.~L., {Vaillancourt}, J.~E., \&
  {Hildebrand}, R.~H. 2016, \apj, 820, 38, \dodoi{10.3847/0004-637X/820/1/38}

\bibitem[{{Houde} {et~al.}(2011){Houde}, {Rao}, {Vaillancourt}, \&
  {Hildebrand}}]{Hou11}
{Houde}, M., {Rao}, R., {Vaillancourt}, J.~E., \& {Hildebrand}, R.~H. 2011,
  \apj, 733, 109, \dodoi{10.1088/0004-637X/733/2/109}

\bibitem[{{Houde} {et~al.}(2009){Houde}, {Vaillancourt}, {Hildebrand},
  {Chitsazzadeh}, \& {Kirby}}]{Hou09}
{Houde}, M., {Vaillancourt}, J.~E., {Hildebrand}, R.~H., {Chitsazzadeh}, S., \&
  {Kirby}, L. 2009, \apj, 706, 1504, \dodoi{10.1088/0004-637X/706/2/1504}

\bibitem[{{Hull} {et~al.}(2014){Hull}, {Plambeck}, {Kwon}, {Bower},
  {Carpenter}, {Crutcher}, {Fiege}, {Franzmann}, {Hakobian}, {Heiles}, {Houde},
  {Hughes}, {Lamb}, {Looney}, {Marrone}, {Matthews}, {Pillai}, {Pound},
  {Rahman}, {Sandell}, {Stephens}, {Tobin}, {Vaillancourt}, {Volgenau}, \&
  {Wright}}]{Hull14}
{Hull}, C. L.~H., {Plambeck}, R.~L., {Kwon}, W., {et~al.} 2014, \apjs, 213, 13,
  \dodoi{10.1088/0067-0049/213/1/13}

\bibitem[{{Hull} {et~al.}(2017){Hull}, {Mocz}, {Burkhart}, {Goodman}, {Girart},
  {Cort{\'e}s}, {Hernquist}, {Springel}, {Li}, \& {Lai}}]{Hul17}
{Hull}, C. L.~H., {Mocz}, P., {Burkhart}, B., {et~al.} 2017, \apjl, 842, L9,
  \dodoi{10.3847/2041-8213/aa71b7}

\bibitem[{Hunter(2007)}]{matplotlib}
Hunter, J.~D. 2007, Computing in Science and Engineering, 9, 90,
  \dodoi{10.1109/MCSE.2007.55}

\bibitem[{{Hwang} {et~al.}(2021){Hwang}, {Kim}, {Pattle}, {Kwon}, {Sadavoy},
  {Koch}, {Hull}, {Johnstone}, {Furuya}, {Won Lee}, {Arzoumanian}, {Tahani},
  {Eswaraiah}, {Liu}, {Kirchschlager}, {Kim}, {Tamura}, {Kwon}, {Lyo}, {Soam},
  {Kang}, {Bourke}, {Matsumura}, {Mairs}, {Kim}, {Park}, {Nakamura}, {Onaka},
  {Tang}, {Liu}, {Ward-Thompson}, {Li}, {Hoang}, {Hasegawa}, {Qiu}, {Lai}, \&
  {Bastien}}]{Hwa21}
{Hwang}, J., {Kim}, J., {Pattle}, K., {et~al.} 2021, \apj, 913, 85,
  \dodoi{10.3847/1538-4357/abf3c4}

\bibitem[{{Ib{\'a}{\~n}ez-Mej{\'\i}a}
  {et~al.}(2022){Ib{\'a}{\~n}ez-Mej{\'\i}a}, {Mac Low}, \& {Klessen}}]{Iba22}
{Ib{\'a}{\~n}ez-Mej{\'\i}a}, J.~C., {Mac Low}, M.-M., \& {Klessen}, R.~S. 2022,
  \apj, 925, 196, \dodoi{10.3847/1538-4357/ac3b58}

\bibitem[{{Inoue} {et~al.}(2018){Inoue}, {Hennebelle}, {Fukui}, {Matsumoto},
  {Iwasaki}, \& {Inutsuka}}]{Ino18}
{Inoue}, T., {Hennebelle}, P., {Fukui}, Y., {et~al.} 2018, \pasj, 70, S53,
  \dodoi{10.1093/pasj/psx089}

\bibitem[{{Joos} {et~al.}(2012){Joos}, {Hennebelle}, \& {Ciardi}}]{Joo12}
{Joos}, M., {Hennebelle}, P., \& {Ciardi}, A. 2012, \aap, 543, A128,
  \dodoi{10.1051/0004-6361/201118730}

\bibitem[{{Kim} {et~al.}(2021){Kim}, {Ostriker}, \& {Filippova}}]{Kim21}
{Kim}, J.-G., {Ostriker}, E.~C., \& {Filippova}, N. 2021, \apj, 911, 128,
  \dodoi{10.3847/1538-4357/abe934}

\bibitem[{{Koch} {et~al.}(2014){Koch}, {Tang}, {Ho}, {Zhang}, {Girart}, {Chen},
  {Frau}, {Li}, {Li}, {Liu}, {Padovani}, {Qiu}, {Yen}, {Chen}, {Ching}, {Lai},
  \& {Rao}}]{Koc14}
{Koch}, P.~M., {Tang}, Y.-W., {Ho}, P. T.~P., {et~al.} 2014, \apj, 797, 99,
  \dodoi{10.1088/0004-637X/797/2/99}

\bibitem[{{Kwon} {et~al.}(2018){Kwon}, {Doi}, {Tamura}, {Matsumura}, {Pattle},
  {Berry}, {Sadavoy}, {Matthews}, {Ward-Thompson}, {Hasegawa}, {Furuya}, {Pon},
  {Di Francesco}, {Arzoumanian}, {Hayashi}, {Kawabata}, {Onaka}, {Choi},
  {Kang}, {Hoang}, {Lee}, {Lee}, {Liu}, {Liu}, {Inutsuka}, {Eswaraiah},
  {Bastien}, {Kwon}, {Lai}, {Qiu}, {Coud{\'e}}, {Franzmann}, {Friberg},
  {Graves}, {Greaves}, {Houde}, {Johnstone}, {Kirk}, {Koch}, {Li}, {Parsons},
  {Rao}, {Rawlings}, {Shinnaga}, {van Loo}, {Aso}, {Byun}, {Chen}, {Chen},
  {Chen}, {Ching}, {Cho}, {Chrysostomou}, {Chung}, {Drabek-Maunder}, {Eyres},
  {Fiege}, {Friesen}, {Fuller}, {Gledhill}, {Griffin}, {Gu}, {Hatchell},
  {Holland}, {Inoue}, {Iwasaki}, {Jeong}, {Kang}, {Kang}, {Kemper}, {Kim},
  {Kim}, {Kim}, {Kim}, {Kim}, {Kim}, {Lacaille}, {Lee}, {Li}, {Li}, {Liu},
  {Liu}, {Lyo}, {Mairs}, {Moriarty-Schieven}, {Nakamura}, {Nakanishi},
  {Ohashi}, {Peretto}, {Pyo}, {Qian}, {Retter}, {Richer}, {Rigby},
  {Robitaille}, {Savini}, {Scaife}, {Soam}, {Tang}, {Tomisaka}, {Wang}, {Wang},
  {Whitworth}, {Yen}, {Yoo}, {Yuan}, {Zhang}, {Zhang}, {Zhou}, {Zhu},
  {Andr{\'e}}, {Dowell}, {Falle}, {Tsukamoto}, {Nakagawa}, {Kanamori},
  {Kataoka}, {Kobayashi}, {Nagata}, {Saito}, {Seta}, \& {Zenko}}]{Kwo18}
{Kwon}, J., {Doi}, Y., {Tamura}, M., {et~al.} 2018, \apj, 859, 4,
  \dodoi{10.3847/1538-4357/aabd82}

\bibitem[{{Lazarian} \& {Hoang}(2007)}]{LazHoa07}
{Lazarian}, A., \& {Hoang}, T. 2007, \mnras, 378, 910,
  \dodoi{10.1111/j.1365-2966.2007.11817.x}

\bibitem[{{Lazarian} {et~al.}(2022){Lazarian}, {Yuen}, \& {Pogosyan}}]{Laz22}
{Lazarian}, A., {Yuen}, K.~H., \& {Pogosyan}, D. 2022, \apj, 935, 77,
  \dodoi{10.3847/1538-4357/ac6877}

\bibitem[{{Lee} {et~al.}(2021){Lee}, {Berthoud}, {Chen}, {Cox}, {Davidson},
  {Encalada}, {Fissel}, {Harrison}, {Kwon}, {Li}, {Li}, {Looney}, {Novak},
  {Sadavoy}, {Santos}, {Segura-Cox}, \& {Stephens}}]{Lee21}
{Lee}, D., {Berthoud}, M., {Chen}, C.-Y., {et~al.} 2021, \apj, 918, 39,
  \dodoi{10.3847/1538-4357/ac0cf2}

\bibitem[{{Li} {et~al.}(2022){Li}, {Lopez-Rodriguez}, {Ajeddig}, {Andr{\'e}},
  {McKee}, {Rho}, \& {Klein}}]{Li22}
{Li}, P.~S., {Lopez-Rodriguez}, E., {Ajeddig}, H., {et~al.} 2022, \mnras, 510,
  6085, \dodoi{10.1093/mnras/stab3448}

\bibitem[{{Liu} {et~al.}(2021){Liu}, {Zhang}, {Commer{\c{c}}on}, {Valdivia},
  {Maury}, \& {Qiu}}]{Liu21}
{Liu}, J., {Zhang}, Q., {Commer{\c{c}}on}, B., {et~al.} 2021, \apj, 919, 79,
  \dodoi{10.3847/1538-4357/ac0cec}

\bibitem[{{Liu} {et~al.}(2022){Liu}, {Zhang}, \& {Qiu}}]{Liu22}
{Liu}, J., {Zhang}, Q., \& {Qiu}, K. 2022, Frontiers in Astronomy and Space
  Sciences, 9, 943556, \dodoi{10.3389/fspas.2022.943556}

\bibitem[{{Lomax} {et~al.}(2015){Lomax}, {Whitworth}, \& {Hubber}}]{Lom15}
{Lomax}, O., {Whitworth}, A.~P., \& {Hubber}, D.~A. 2015, \mnras, 449, 662,
  \dodoi{10.1093/mnras/stv310}

\bibitem[{{Menten} {et~al.}(2007){Menten}, {Reid}, {Forbrich}, \&
  {Brunthaler}}]{Men07}
{Menten}, K.~M., {Reid}, M.~J., {Forbrich}, J., \& {Brunthaler}, A. 2007, \aap,
  474, 515, \dodoi{10.1051/0004-6361:20078247}

\bibitem[{{Mestel} \& {Spitzer}(1956)}]{Mes56}
{Mestel}, L., \& {Spitzer}, L., J. 1956, \mnras, 116, 503,
  \dodoi{10.1093/mnras/116.5.503}

\bibitem[{{Myers} \& {Goodman}(1991)}]{MyeGoo91}
{Myers}, P.~C., \& {Goodman}, A.~A. 1991, \apj, 373, 509,
  \dodoi{10.1086/170070}

\bibitem[{{Ngoc} {et~al.}(2023){Ngoc}, {Diep}, {Hoang}, {Tram}, {Giang},
  {L{\^e}}, {Hoang}, {Phuong}, {Khang}, {Nguyen}, \& {Truong}}]{Ngoc23}
{Ngoc}, N.~B., {Diep}, P.~N., {Hoang}, T., {et~al.} 2023, \apj, 953, 66,
  \dodoi{10.3847/1538-4357/acdb6e}

\bibitem[{{Ostriker} {et~al.}(2001){Ostriker}, {Stone}, \& {Gammie}}]{Ost01}
{Ostriker}, E.~C., {Stone}, J.~M., \& {Gammie}, C.~F. 2001, \apj, 546, 980,
  \dodoi{10.1086/318290}

\bibitem[{{Padoan} {et~al.}(2001){Padoan}, {Goodman}, {Draine}, {Juvela},
  {Nordlund}, \& {R{\"o}gnvaldsson}}]{Pad01}
{Padoan}, P., {Goodman}, A., {Draine}, B.~T., {et~al.} 2001, \apj, 559, 1005,
  \dodoi{10.1086/322504}

\bibitem[{{Pattle} \& {Fissel}(2019)}]{PatFis19}
{Pattle}, K., \& {Fissel}, L. 2019, Frontiers in Astronomy and Space Sciences,
  6, 15, \dodoi{10.3389/fspas.2019.00015}

\bibitem[{{Pattle} {et~al.}(2023){Pattle}, {Fissel}, {Tahani}, {Liu}, \&
  {Ntormousi}}]{Pat23}
{Pattle}, K., {Fissel}, L., {Tahani}, M., {Liu}, T., \& {Ntormousi}, E. 2023,
  in Astronomical Society of the Pacific Conference Series, Vol. 534,
  Astronomical Society of the Pacific Conference Series, ed. S.~{Inutsuka},
  Y.~{Aikawa}, T.~{Muto}, K.~{Tomida}, \& M.~{Tamura}, 193

\bibitem[{{Pattle} {et~al.}(2017){Pattle}, {Ward-Thompson}, {Berry},
  {Hatchell}, {Chen}, {Pon}, {Koch}, {Kwon}, {Kim}, {Bastien}, {Cho},
  {Coud{\'e}}, {Di Francesco}, {Fuller}, {Furuya}, {Graves}, {Johnstone},
  {Kirk}, {Kwon}, {Lee}, {Matthews}, {Mottram}, {Parsons}, {Sadavoy},
  {Shinnaga}, {Soam}, {Hasegawa}, {Lai}, {Qiu}, \& {Friberg}}]{Pat17}
{Pattle}, K., {Ward-Thompson}, D., {Berry}, D., {et~al.} 2017, \apj, 846, 122,
  \dodoi{10.3847/1538-4357/aa80e5}

\bibitem[{{Pillai} {et~al.}(2020){Pillai}, {Clemens}, {Reissl}, {Myers},
  {Kauffmann}, {Lopez-Rodriguez}, {Alves}, {Franco}, {Henshaw}, {Menten},
  {Nakamura}, {Seifried}, {Sugitani}, \& {Wiesemeyer}}]{Pil20}
{Pillai}, T. G.~S., {Clemens}, D.~P., {Reissl}, S., {et~al.} 2020, Nature
  Astronomy, 4, 1195, \dodoi{10.1038/s41550-020-1172-6}

\bibitem[{{Pineda} {et~al.}(2023){Pineda}, {Arzoumanian}, {Andre}, {Friesen},
  {Zavagno}, {Clarke}, {Inoue}, {Chen}, {Lee}, {Soler}, \& {Kuffmeier}}]{Pin23}
{Pineda}, J.~E., {Arzoumanian}, D., {Andre}, P., {et~al.} 2023, in Astronomical
  Society of the Pacific Conference Series, Vol. 534, Astronomical Society of
  the Pacific Conference Series, ed. S.~{Inutsuka}, Y.~{Aikawa}, T.~{Muto},
  K.~{Tomida}, \& M.~{Tamura}, 233

\bibitem[{{Planck Collaboration} {et~al.}(2020){Planck Collaboration},
  {Aghanim}, {Akrami}, {Alves}, {Ashdown}, {Aumont}, {Baccigalupi},
  {Ballardini}, {Banday}, {Barreiro}, {Bartolo}, {Basak}, {Benabed}, {Bernard},
  {Bersanelli}, {Bielewicz}, {Bock}, {Bond}, {Borrill}, {Bouchet}, {Boulanger},
  {Bracco}, {Bucher}, {Burigana}, {Calabrese}, {Cardoso}, {Carron}, {Chary},
  {Chiang}, {Colombo}, {Combet}, {Crill}, {Cuttaia}, {de Bernardis}, {de
  Zotti}, {Delabrouille}, {Delouis}, {Di Valentino}, {Dickinson}, {Diego},
  {Dor{\'e}}, {Douspis}, {Ducout}, {Dupac}, {Efstathiou}, {Elsner},
  {En{\ss}lin}, {Eriksen}, {Falgarone}, {Fantaye}, {Fernandez-Cobos},
  {Ferri{\`e}re}, {Finelli}, {Forastieri}, {Frailis}, {Fraisse}, {Franceschi},
  {Frolov}, {Galeotta}, {Galli}, {Ganga}, {G{\'e}nova-Santos}, {Gerbino},
  {Ghosh}, {Gonz{\'a}lez-Nuevo}, {G{\'o}rski}, {Gratton}, {Green}, {Gruppuso},
  {Gudmundsson}, {Guillet}, {Handley}, {Hansen}, {Helou}, {Herranz}, {Hivon},
  {Huang}, {Jaffe}, {Jones}, {Keih{\"a}nen}, {Keskitalo}, {Kiiveri}, {Kim},
  {Krachmalnicoff}, {Kunz}, {Kurki-Suonio}, {Lagache}, {Lamarre}, {Lasenby},
  {Lattanzi}, {Lawrence}, {Le Jeune}, {Levrier}, {Liguori}, {Lilje},
  {Lindholm}, {L{\'o}pez-Caniego}, {Lubin}, {Ma}, {Mac{\'\i}as-P{\'e}rez},
  {Maggio}, {Maino}, {Mandolesi}, {Mangilli}, {Marcos-Caballero}, {Maris},
  {Martin}, {Mart{\'\i}nez-Gonz{\'a}lez}, {Matarrese}, {Mauri}, {McEwen},
  {Melchiorri}, {Mennella}, {Migliaccio}, {Miville-Desch{\^e}nes}, {Molinari},
  {Moneti}, {Montier}, {Morgante}, {Moss}, {Natoli}, {Pagano}, {Paoletti},
  {Patanchon}, {Perrotta}, {Pettorino}, {Piacentini}, {Polastri}, {Polenta},
  {Puget}, {Rachen}, {Reinecke}, {Remazeilles}, {Renzi}, {Ristorcelli},
  {Rocha}, {Rosset}, {Roudier}, {Rubi{\~n}o-Mart{\'\i}n}, {Ruiz-Granados},
  {Salvati}, {Sandri}, {Savelainen}, {Scott}, {Sirignano}, {Sunyaev},
  {Suur-Uski}, {Tauber}, {Tavagnacco}, {Tenti}, {Toffolatti}, {Tomasi},
  {Trombetti}, {Valiviita}, {Vansyngel}, {Van Tent}, {Vielva}, {Villa},
  {Vittorio}, {Wandelt}, {Wehus}, {Zacchei}, \& {Zonca}}]{PlanckXII}
{Planck Collaboration}, {Aghanim}, N., {Akrami}, Y., {et~al.} 2020, \aap, 641,
  A12, \dodoi{10.1051/0004-6361/201833885}

\bibitem[{{Schive} {et~al.}(2018){Schive}, {ZuHone}, {Goldbaum}, {Turk},
  {Gaspari}, \& {Cheng}}]{Sch18}
{Schive}, H.-Y., {ZuHone}, J.~A., {Goldbaum}, N.~J., {et~al.} 2018, \mnras,
  481, 4815, \dodoi{10.1093/mnras/sty2586}

\bibitem[{{Seifried} {et~al.}(2011){Seifried}, {Banerjee}, {Klessen}, {Duffin},
  \& {Pudritz}}]{Sei11}
{Seifried}, D., {Banerjee}, R., {Klessen}, R.~S., {Duffin}, D., \& {Pudritz},
  R.~E. 2011, \mnras, 417, 1054, \dodoi{10.1111/j.1365-2966.2011.19320.x}

\bibitem[{{Seifried} {et~al.}(2020){Seifried}, {Haid}, {Walch}, {Borchert}, \&
  {Bisbas}}]{Sei20}
{Seifried}, D., {Haid}, S., {Walch}, S., {Borchert}, E.~M.~A., \& {Bisbas},
  T.~G. 2020, \mnras, 492, 1465, \dodoi{10.1093/mnras/stz3563}

\bibitem[{{Skalidis} {et~al.}(2021){Skalidis}, {Sternberg}, {Beattie},
  {Pavlidou}, \& {Tassis}}]{Ska21}
{Skalidis}, R., {Sternberg}, J., {Beattie}, J.~R., {Pavlidou}, V., \& {Tassis},
  K. 2021, \aap, 656, A118, \dodoi{10.1051/0004-6361/202142045}

\bibitem[{{Tang} {et~al.}(2019){Tang}, {Koch}, {Peretto}, {Novak},
  {Duarte-Cabral}, {Chapman}, {Hsieh}, \& {Yen}}]{Tan19}
{Tang}, Y.-W., {Koch}, P.~M., {Peretto}, N., {et~al.} 2019, \apj, 878, 10,
  \dodoi{10.3847/1538-4357/ab1484}

\bibitem[{Virtanen {et~al.}(2020)Virtanen, Gommers, Oliphant, Haberland, Reddy,
  Cournapeau, Burovski, Peterson, Weckesser, Bright, {van der Walt}, Brett,
  Wilson, Millman, Mayorov, Nelson, Jones, Kern, Larson, Carey, Polat, Feng,
  Moore, {VanderPlas}, Laxalde, Perktold, Cimrman, Henriksen, Quintero, Harris,
  Archibald, Ribeiro, Pedregosa, {van Mulbregt}, \& {SciPy 1.0
  Contributors}}]{scipy}
Virtanen, P., Gommers, R., Oliphant, T.~E., {et~al.} 2020, Nature Methods, 17,
  261, \dodoi{10.1038/s41592-019-0686-2}

\bibitem[{{Wang} {et~al.}(2019){Wang}, {Lai}, {Eswaraiah}, {Pattle}, {Di
  Francesco}, {Johnstone}, {Koch}, {Liu}, {Tamura}, {Furuya}, {Onaka},
  {Ward-Thompson}, {Soam}, {Kim}, {Lee}, {Lee}, {Mairs}, {Arzoumanian}, {Kim},
  {Hoang}, {Hwang}, {Liu}, {Berry}, {Bastien}, {Hasegawa}, {Kwon}, {Qiu},
  {Andr{\'e}}, {Aso}, {Byun}, {Chen}, {Chen}, {Chen}, {Ching}, {Cho}, {Choi},
  {Chrysostomou}, {Chung}, {Coud{\'e}}, {Doi}, {Dowell}, {Drabek-Maunder},
  {Duan}, {Eyres}, {Falle}, {Fanciullo}, {Fiege}, {Franzmann}, {Friberg},
  {Friesen}, {Fuller}, {Gledhill}, {Graves}, {Greaves}, {Griffin}, {Gu}, {Han},
  {Hatchell}, {Hayashi}, {Holland}, {Houde}, {Inoue}, {Inutsuka}, {Iwasaki},
  {Jeong}, {Kanamori}, {Kang}, {Kang}, {Kang}, {Kataoka}, {Kawabata}, {Kemper},
  {Kim}, {Kim}, {Kim}, {Kim}, {Kirk}, {Kobayashi}, {Konyves}, {Kwon},
  {Lacaille}, {Lee}, {Lee}, {Lee}, {Lee}, {Li}, {Li}, {Li}, {Liu}, {Liu},
  {Lyo}, {Matsumura}, {Matthews}, {Moriarty-Schieven}, {Nagata}, {Nakamura},
  {Nakanishi}, {Ohashi}, {Park}, {Parsons}, {Pascale}, {Peretto}, {Pon}, {Pyo},
  {Qian}, {Rao}, {Rawlings}, {Retter}, {Richer}, {Rigby}, {Robitaille},
  {Sadavoy}, {Saito}, {Savini}, {Scaife}, {Seta}, {Shinnaga}, {Tang},
  {Tomisaka}, {Tsukamoto}, {van Loo}, {Wang}, {Whitworth}, {Yen}, {Yoo},
  {Yuan}, {Yun}, {Zenko}, {Zhang}, {Zhang}, {Zhang}, {Zhou}, \& {Zhu}}]{Wan19}
{Wang}, J.-W., {Lai}, S.-P., {Eswaraiah}, C., {et~al.} 2019, \apj, 876, 42,
  \dodoi{10.3847/1538-4357/ab13a2}

\bibitem[{{Wang} {et~al.}(2024){Wang}, {Koch}, {Clarke}, {Fuller}, {Peretto},
  {Tang}, {Yen}, {Lai}, {Ohashi}, {Arzoumanian}, {Johnstone}, {Furuya},
  {Inutsuka}, {Lee}, {Ward-Thompson}, {Le Gouellec}, {Liu}, {Fanciullo},
  {Hwang}, {Pattle}, {Poidevin}, {Tahani}, {Onaka}, {Rawlings}, {Chung}, {Liu},
  {Lyo}, {Priestley}, {Hoang}, {Tamura}, {Berry}, {Bastien}, {Ching},
  {Coud{\'e}}, {Kwon}, {Chen}, {Eswaraiah}, {Soam}, {Hasegawa}, {Qiu},
  {Bourke}, {Byun}, {Chen}, {Chen}, {Chen}, {Cho}, {Choi}, {Choi}, {Choi},
  {Chrysostomou}, {Dai}, {Di Francesco}, {Diep}, {Doi}, {Duan}, {Duan}, {Eden},
  {Fiege}, {Fissel}, {Franzmann}, {Friberg}, {Friesen}, {Gledhill}, {Graves},
  {Greaves}, {Griffin}, {Gu}, {Han}, {Hayashi}, {Houde}, {Inoue}, {Iwasaki},
  {Jeong}, {K{\"o}nyves}, {Kang}, {Kang}, {Karoly}, {Kataoka}, {Kawabata},
  {Khan}, {Kim}, {Kim}, {Kim}, {Kim}, {Kim}, {Kim}, {Kim}, {Kirchschlager},
  {Kirk}, {Kobayashi}, {Kusune}, {Kwon}, {Lacaille}, {Law}, {Lee}, {Lee},
  {Lee}, {Lee}, {Li}, {Li}, {Li}, {Li}, {Lin}, {Liu}, {Liu}, {Lu}, {Mairs},
  {Matsumura}, {Matthews}, {Moriarty-Schieven}, {Nagata}, {Nakamura},
  {Nakanishi}, {Ngoc}, {Park}, {Parsons}, {Pyo}, {Qian}, {Rao}, {Rawlings},
  {Retter}, {Richer}, {Rigby}, {Sadavoy}, {Saito}, {Savini}, {Seta}, {Sharma},
  {Shimajiri}, {Shinnaga}, {Tang}, {Thuong}, {Tomisaka}, {Tram}, {Tsukamoto},
  {Viti}, {Wang}, {Whitworth}, {Wu}, {Xie}, {Yang}, {Yoo}, {Yuan}, {Yun},
  {Zenko}, {Zhang}, {Zhang}, {Zhang}, {Zhou}, {Zhu}, {de Looze}, {Andr{\'e}},
  {Dowell}, {Eyres}, {Falle}, {Robitaille}, \& {van Loo}}]{Wang24}
{Wang}, J.-W., {Koch}, P.~M., {Clarke}, S.~D., {et~al.} 2024, \apj, 962, 136,
  \dodoi{10.3847/1538-4357/ad165b}

\bibitem[{{Ward-Thompson} {et~al.}(2017){Ward-Thompson}, {Pattle}, {Bastien},
  {Furuya}, {Kwon}, {Lai}, {Qiu}, {Berry}, {Choi}, {Coud{\'e}}, {Di Francesco},
  {Hoang}, {Franzmann}, {Friberg}, {Graves}, {Greaves}, {Houde}, {Johnstone},
  {Kirk}, {Koch}, {Kwon}, {Lee}, {Li}, {Matthews}, {Mottram}, {Parsons}, {Pon},
  {Rao}, {Rawlings}, {Shinnaga}, {Sadavoy}, {van Loo}, {Aso}, {Byun},
  {Eswaraiah}, {Chen}, {Chen}, {Chen}, {Ching}, {Cho}, {Chrysostomou}, {Chung},
  {Doi}, {Drabek-Maunder}, {Eyres}, {Fiege}, {Friesen}, {Fuller}, {Gledhill},
  {Griffin}, {Gu}, {Hasegawa}, {Hatchell}, {Hayashi}, {Holland}, {Inoue},
  {Inutsuka}, {Iwasaki}, {Jeong}, {Kang}, {Kang}, {Kang}, {Kawabata}, {Kemper},
  {Kim}, {Kim}, {Kim}, {Kim}, {Kim}, {Kim}, {Lacaille}, {Lee}, {Lee}, {Li},
  {Li}, {Liu}, {Liu}, {Liu}, {Liu}, {Lyo}, {Mairs}, {Matsumura},
  {Moriarty-Schieven}, {Nakamura}, {Nakanishi}, {Ohashi}, {Onaka}, {Peretto},
  {Pyo}, {Qian}, {Retter}, {Richer}, {Rigby}, {Robitaille}, {Savini}, {Scaife},
  {Soam}, {Tamura}, {Tang}, {Tomisaka}, {Wang}, {Wang}, {Whitworth}, {Yen},
  {Yoo}, {Yuan}, {Zhang}, {Zhang}, {Zhou}, {Zhu}, {Andr{\'e}}, {Dowell},
  {Falle}, \& {Tsukamoto}}]{War17}
{Ward-Thompson}, D., {Pattle}, K., {Bastien}, P., {et~al.} 2017, \apj, 842, 66,
  \dodoi{10.3847/1538-4357/aa70a0}

\bibitem[{{Whittet} {et~al.}(2008){Whittet}, {Hough}, {Lazarian}, \&
  {Hoang}}]{Whi08}
{Whittet}, D.~C.~B., {Hough}, J.~H., {Lazarian}, A., \& {Hoang}, T. 2008, \apj,
  674, 304, \dodoi{10.1086/525040}

\bibitem[{{Wiebe} \& {Watson}(2004)}]{WieWat04}
{Wiebe}, D.~S., \& {Watson}, W.~D. 2004, \apj, 615, 300, \dodoi{10.1086/424033}

\bibitem[{{Wurster} {et~al.}(2019){Wurster}, {Bate}, \& {Price}}]{Wur19}
{Wurster}, J., {Bate}, M.~R., \& {Price}, D.~J. 2019, \mnras, 489, 1719,
  \dodoi{10.1093/mnras/stz2215}

\bibitem[{{Wurster} {et~al.}(2016){Wurster}, {Price}, \& {Bate}}]{Wur16}
{Wurster}, J., {Price}, D.~J., \& {Bate}, M.~R. 2016, \mnras, 457, 1037,
  \dodoi{10.1093/mnras/stw013}

\end{thebibliography}
\bibliographystyle{aasjournal}



\end{document}